\documentclass[aps,pra,twocolumn,showpacs,superscriptaddress,groupedaddress]{revtex4-1}

\pdfoutput=1

\usepackage[utf8]{inputenc}
\usepackage[T1]{fontenc}
\usepackage{graphicx}
\usepackage{latexsym,amsmath,amssymb,amsthm}
\usepackage{subfigure}
\usepackage{float}
\usepackage{textcomp}
\usepackage{url}
\usepackage{sidecap}
\usepackage{color}
\usepackage{times}

\newcommand{\kk}{\mathbf{k}}
\newcommand{\qq}{\mathbf{q}}
\newcommand{\xx}{\mathbf{x}}

\newcommand{\dd}{\mathrm{d}}
\newcommand{\ee}{\mathrm{e}}

\newcommand{\dsp}{\delta S^+_\kk}
\newcommand{\dsm}{\delta S^-_\kk}

\newcommand{\dg}{^\dagger}
\newcommand{\bh}{{\rm{b}}}
\newcommand{\bhd}{{\rm b}^\dagger}

\mathchardef\mhyphen="2D

\begin{document}

\title{A quantum Langevin model for non-equilibrium {condensation}}
\author{Alessio Chiocchetta}
\affiliation{ SISSA - International School for Advanced Studies, via Bonomea 265, 34136 Trieste, Italy \\ INFN - Istituto Nazionale di Fisica Nucleare, Sezione di Trieste, Italy}
\author{Iacopo Carusotto}
\affiliation{INO-CNR BEC Center and Dipartimento di Fisica, Universit\`a di Trento, Via Sommarive 14, I-38123 Povo, Italy}

\date{\today}

\begin{abstract}
{We develop a quantum model for non-equilibrium Bose-Einstein condensation of photons and polaritons in planar microcavity devices. The model builds upon laser theory and includes the spatial dynamics of the cavity field, a saturation mechanism and some frequency-dependence of the gain: quantum Langevin equations are written for a cavity field coupled to a continuous distribution of externally pumped two-level emitters with a well-defined frequency. As a an example of application, the method is used to study the linearised quantum fluctuations around a steady-state condensed state. In the good-cavity regime, an effective equation for the cavity field only is proposed in terms of a stochastic Gross-Pitaevskii equation. Perspectives in view of a full quantum simulation of the non-equilibrium condensation process are finally sketched.}
\end{abstract}

\maketitle

\section{Introduction}
\label{sec:Introduction}

Recent experimental demonstrations of Bose-Einstein condensation (BEC) phenomena in luminous gases of exciton-polaritons{~\cite{STEVENSON2000,BAUMBERG2000,BAAS2006,KASP2006}} and pure photons~\cite{KLAE2010} in optical microcavities are opening exciting new perspectives to the study of non-equilibrium statistical mechanics of open, driven-dissipative systems. In contrast to usual statistical mechanics where the equilibrium density matrix is determined by the Boltzmann factor $\rho_{\rm eq}\propto \exp(-H/k_B T)$, the steady-state of open systems is determined by a dynamical balance of pumping and losses. The novel features that stem from this difference are presently attracting a lot of interest from both theoretical and experimental points of view, in particular for what concerns phase transitions and critical behavior~\cite{Zia, Cross, Henkel}. 

In optics, the first and most celebrated example of phase transition is the laser operation threshold and its interpretation in terms of a spontaneously broken $U(1)$ phase symmetry {was} first pointed out in the early 1970's~\cite{Graham,DEGI1970,HAKE1975}. While this {analogy with Bose-Einstein condensation (BEC)} is typically discussed in textbooks for the case of single-mode laser cavities, {rigorously speaking} the concepts of phase transition and of spontaneous symmetry-breaking phenomenon are restricted to spatially infinite systems. {Only recently, the} advances in optical technology are providing examples of spatially extended laser devices for which the large system limit is a legitimate approximation, the so-called VCSELs (Vertical Cavity Surface Emitting Lasers)~\cite{VCSELbook}. While these devices have received a great attention from point of view of nonlinear optics and of all-optical information processing~\cite{VCSEL}, their potential to study the non-equilibrium statistical mechanics of the laser phase transition has been so far only marginally exploited~\cite{KAPON}.

As it is reviewed in~\cite{RMP2013}, the interest {for these condensation} phenomena in optical systems got strongly revived in the last decade with the experimental observations of polariton and photon BECs~\cite{STEVENSON2000,BAUMBERG2000,KASP2006,KLAE2010}. {As a remarkable difference from standard lasers, it was pointed out that the effective interactions between the individual particles forming the photon and polariton gases mediated by the underlying medium may lead to collective behaviours in the gas including, e.g., superfluidity~\cite{WOUT2010}. 

At the same time, a significant work has been devoted to characterize the equilibrium vs. non-equilibrium nature of these condensates and quantify the observable consequences of the pumping and loss processes. On one hand, the photon BEC experiment of \cite{KLAE2010} has shown clear evidence of a thermal Bose-Einstein distribution at the temperature of the cavity medium embedding the dye molecules. On the other hand, qualitatively novel features of non-equilibrium BEC have been observed in polariton condensation experiments. For example, the early experiments of \cite{Richard:PRL2005} have shown BEC into a ring of modes at finite $\kk$: An interpretation of this effect in terms of an interplay of driving, dissipation and energy minimization was proposed in \cite{WOUT2008} and experimentally confirmed by \cite{WERTZ}. Another, even more surprising feature was experimentally reported in \cite{BAJO2007}, where a thermal-like distribution was observed even in a weak coupling regime where collisions are expected to be too weak to allow for any thermalisation.

From the theoretical point of view, the recent work~\cite{KEELING2013} has quantitatively explored the crossover from the equilibrium-like regime of~\cite{KLAE2010} where the particle distribution closely follow the Bose-Einstein distribution, to non-equilibrium regimes where the distribution is more and more distorted up to the standard laser regime: in particular, the ratio between the thermalisation rate (encoded by the absorption/emission rates) and the pumping and photon losses was identified as the key parameter determining the equilibrium vs. non-equilibrium nature of the momentum distribution of photons.

Going beyond the one-body distribution function, several authors~\cite{SZYM2006, WOUT2006,WOUT2007} have pointed out a qualitative signature of non-equilibrium in the dispersion of the collective excitations: the typical acoustic branch of equilibrium condensates is replaced by a diffusive plateau at low wavevectors, whose $\kk$-space extension is quantitatively related to the departure from equilibrium. Furthermore,} the non-perturbative functional renormalisation group calculation in~\cite{Sieberer2013} showed the importance of new critical exponents arising from the genuine non-equilibrium nature of the system. Finally, theoretical descriptions of the photon BEC phenomenon in purely laser terms were {aimed for in}~\cite{Fischer}. {An interesting proposal to obtain a chemical potential for photons was proposed in~\cite{hafezi2014chemical}.}

{The situation is even more intriguing in the reduced dimension case that is naturally realized in experiments:} While a well-developed condensate with spatial coherence extending in the whole gas was observed in the relatively small systems of~\cite{KASP2006,KLAE2010}, quasi-condensation features are expected to arise in larger systems {because of long-wavelength fluctuations. In the equilibrium case, the well-known Mermin-Wagner theorem forbids BEC in {translationally invariant systems \footnote{{Of course, BEC remains possible if a harmonic trap potential is added to the 2D gas~\cite{PITA2004}, as done in the experiment of~\cite{KLAE2010}}} of dimension smaller or equal to 2~\cite{MERM1966}. In the non-equilibrium case,} first theoretical works based on a Gaussian linearised theory of fluctuations have anticipated that the long-distance behaviour of the non-equilibrium (interacting) quasi-condensate {should be} the same as in the corresponding equilibrium system at finite $T$, that is an exponential decay of coherence in one dimension and a power-law decay in two dimensions~\cite{SZYM2006, WOUT2006, ACIC2013}. Pioneering experiments along these lines were reported in~\cite{SPAN2012,ROUM2012}. 
Very recently, more refined theoretical studies going beyond the Gaussian theory have started questioning some aspects of these theoretical predictions. {In particular,} it was pointed out in~\cite{Altman,Wouters2013} that terms beyond the linearised Bogoliubov theory are essential to correctly capture the long-distance behavior of the spatial coherence {and correct some pathologies found in the non-interacting limit in~\cite{ACIC2013}. As a result,} the power-law quasi-long range order of {spatially homogeneous} two-dimensional quasi-condensates might be broken and replaced by a stretched exponential decay~\cite{Altman}. 

{The common starting point of all these theoretical works are phenomenological stochastic Gross-Pitaevskii equations (SGPE). The only exception is the numerical simulation reported in~\cite{Ciuti2005} where the BEC phase transition was studied in the so-called Optical Parametric Oscillator (OPO) configuration {which is} amenable to an almost {\em ab initio} truncated-Wigner description of the field dynamics. In all other cases, the strength and the functional form of the noise terms had to be introduced in a phenomenological way~\cite{WOUT2009,ACIC2013}.} The purpose of this work is to develop a fully quantum model of the system from which one can derive a SGPE under controlled approximations. In contrast to previous derivations of the SGPE based, e.g., on Keldysh formalism~\cite{Keeling} or on the truncated-Wigner representations of the field~\cite{Ciuti2005,WOUT2009}, our derivation is performed through the quantum Langevin approach~\cite{GARD2000}: on one hand, this approach offers a physically transparent description of the baths and, in particular, of the incoherent pumping mechanism. On the other hand, it allows to capture {within a simple Markovian theory} the frequency-dependence of the pumping and dissipation baths. In the good cavity limit, we can then adiabatically eliminate the matter degrees of freedom, which results in an effective dynamics for the cavity photon field only: in particular, explicit expressions for the Langevin terms are provided, which can eventually be used as a starting point for more sophisticated statistical mechanics calculations.

This Article is organised as follows. In Sec. \ref{sec:QL} we present the model and we derive {the quantum Langevin} equations. In Sec. \ref{sec:MF}, we present the mean-field theory of the condensation process {and we illustrate the $U(1)$ spontaneous symmetry breaking phenomenon}. In the following Sec. \ref{sec:Fluctuations} we study the excitation modes of the system and the effect of fluctuations around the condensate: in particular, predictions for the momentum distribution of the thermal component and for the luminescence spectrum are given. In Sec. \ref{sec:SGPE} we discuss the good cavity limit where our equations can be reduced to a stochastic Gross-Pitaevskii equation. Conclusions are finally drawn in Sec. \ref{sec:Conclusions}.

\section{The model}
\label{sec:QL}
Our microscopic theory extends early models on laser operation~\cite{LAX1966,LOUI1973,GORD1967,HAKE1984} to the spatially extended case of planar cavities with a parabolic dispersion of the cavity photon as a function of the in-plane wavevector $\kk$,
\begin{equation}
\omega_\kk = \omega_0 +  \frac{k^2}{2m}.
\label{omegak}
\end{equation}
with a cut-off frequency $\omega_0$ and an effective mass $m$~\cite{RMP2013}. This simple description of cavity modes well captures the physics of planar DBR semiconductor microcavities {in both the weak and the strong light-matter coupling regimes: in particular, low-momentum polaritons used in the condensation experiment~\cite{KASP2006} are straightforwardly included as dressed photon modes with suitably renormalised $\omega_0$ and $m$ parameters. When} supplemented with an harmonic potential term accounting for the mirror curvature, {this same formalism} also describes the mesoscopic cavity of~\cite{KLAE2010}.

As it is sketched in Fig.\ref{fig:sketch}, the {cavity} field is then coupled to a set of two-level emitters. Both the emitters and the cavity are subject to losses of different natures, while energy is continuously injected into the system by pumping the emitters to their excited state. The steady-state of the system is therefore determined by a dynamical balance of pumping and losses. In this description, {both Bose-Einstein condensation and lasing consist} in the appearance of a macroscopic coherent field in a single mode of the cavity (typically the $\kk=0$ one), monochromatically oscillating at a given frequency $\omega$ and with a long-distance coherence extending in the whole system. Part of the in-cavity light eventually leaves the cavity via the non-perfectly reflecting mirror and end up forming a coherent {output beam of light}. 

\begin{figure}[htbp]
\includegraphics[width=7.5cm]{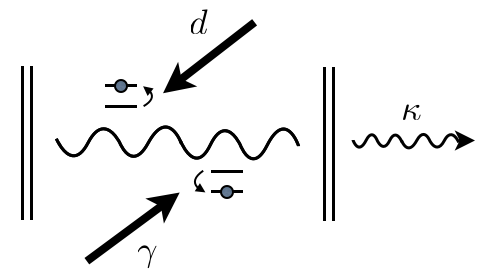}
\caption{\label{fig:sketch}A pictorial representation of the model. {Emitters lose energy at a rate $\gamma$ while energy is pumped in at a rate $d$. Photons can leave the cavity after a time $\kappa^{-1}$.}}
\end{figure}

While this theory directly builds on standard laser theory, it is generic enough to capture the main specificities of exciton-polariton condensation under an incoherent pumping scheme {which was experimentally demonstrated in~\cite{KASP2006}. In this case, the dispersion is the polariton one and the two-level emitters provide a model description of the complex irreversible polariton scattering processes replenishing the condensate~\cite{keeling2007,DENG2010}. The main gain process consists of binary polariton scattering where two polaritons located around the inflection point of their dispersion are scattered into one condensate polariton and one exciton (which is then quickly lost). In our model, the excited state of the emitters correspond to pairs of polaritons located around the inflection point of their dispersion, while the ground state of the emitter corresponds to having one exciton resulting from the collision. At simplest order, the emitter energy $\nu$ is then approximately equal to the difference of the energy of the pair around the inflection point and of the exciton, $\hbar\nu\approx 2E_{\rm infl}-E_{\rm exc}$, that is the energy where the collisional gain is expected to be maximum. }
Extensions of this theory including more complicate emitters can be used to describe the dye molecules involved in the photon condensation experiments of \cite{KLAE2010}. Several possibilities in this direction are explored in \cite{KEELING2013,DELEEUW2013}.

\subsection{The field and emitter Hamiltonians and the radiation-emitter coupling}

Given the translational symmetry of the system along the cavity plane, the in-plane momentum $\kk$ of the photon is a good quantum number and the (bare) photon dispersion of a given longitudinal mode is well described by the  parabolic dispersion \eqref{omegak}. The emitters are fixed in space according to a regular square lattice and do not have any direct interaction.

Taking for notational simplicity $\hbar=1$, the free Hamiltonian of the field and {of} the emitters has the usual form
\begin{equation}
H _{\text{free}} = \sum_\kk \omega_\kk \bhd_\kk \bh_\kk + \sum_i \nu S_i^z 
\end{equation}
where $\omega_\kk$ is the cavity dispersion defined in \eqref{omegak} and $\nu$ is the emitter frequency. The $\bh_\kk,\bhd_\kk$ operators satisfy bosonic commutation rules $[\bh_\kk,\bhd_{\kk'}]=\delta_{\kk,\kk'}$, {while the emitter operators $S^{\pm,z}$ satisfy the usual algebra of spin-1/2 operators.}

Within the usual rotating-wave approximation, the radiation-matter coupling is then:
\begin{equation}
H _{\text{int}} = \frac{ig}{\sqrt{V}}\sum_i\sum_\kk\left( \ee^{i\kk\cdot \xx_i}\,\bh_\kk\,S_{i}^+ - \ee^{-i\kk\cdot \xx_i}\,\bhd_\kk\,S_i^- \right),
\end{equation}
where $\xx_i$ is the position of the $i$-th emitter and $V$ is the total volume of the system. 

Assuming periodic boundary conditions, we can introduce the $D$-dimensional real-space cavity field
\begin{equation}
 \phi(\xx) = \frac{1}{\sqrt{V}}\sum_\kk e^{i\kk\cdot\xx}\,\bh_\kk.
 \end{equation}
In terms of the field $\phi(\xx)$, local binary interactions between the cavity photons can be added to the model via a two-body interaction term of the form
\begin{equation}
 H^{(4)} = \frac{\lambda}{2}\int_V\mathrm{d}^Dx\,\phi^\dagger(\xx)\phi^\dagger(\xx)\phi(\xx)\phi(\xx),
\end{equation}
which in momentum space reads: 
\begin{equation}
 H^{(4)} = \frac{\lambda}{2V}\sum_{\kk\kk'\qq}\bhd_{\kk+\qq} \bhd_{\kk'-\qq} \bh_{\kk'} b_\kk.
\end{equation}
Physically, such a term can describe a Kerr $\chi^{(3)}$ optical non-linearity of the cavity material or, equivalently, polariton-polariton interactions~\cite{RMP2013}.

\subsection{Dissipative field dynamics: radiative losses}

The cavity field is coupled to an external bath of radiative modes via the non-perfectly reflecting cavity mirrors. As usual, this can be modelled by coupling each $\kk$ mode of the field with a bath of harmonic oscillators~\cite{ICCCPRA2006}. The resulting quantum Langevin equations~\cite{GARD2000} then have the form
\begin{equation}
 \frac{d\bhd_\kk}{dt}  =  \left( i\omega_\kk - \frac{\kappa}{2} \right) \bhd_\kk + F_\kk^\dagger.
\end{equation}
Here, $\kappa$ is the decay rate of the field and the zero-mean quantum noises $F_\kk^\dagger$ are uncorrelated and have a delta-like correlation in time
\begin{eqnarray}
  \langle F^\dagger_\kk(t)F_{\kk'}(t')\rangle  &=& 0 \label{F1} \\
 \langle F_\kk(t)F_{\kk'}^\dagger(t')\rangle  &=& \kappa\,\delta(t - t')\,\delta_{\kk,\kk'}. \label{F2}
\end{eqnarray}
This form of the quantum Langevin equation requires that the initial total density matrix factorizes in the cavity and bath parts and that the bath density matrix corresponds to an equilibrium state at very low temperature. Both approximations are well satisfied by realistic systems, since the frequencies involved in optical experiments are very high as compared to the device temperature, typically at or below room temperature. As a result, cavity photons can only spontaneously quit the cavity after a lifetime $\kappa^{-1}$, while no radiation can enter the cavity from outside.

\subsection{Dissipative emitter dynamics: losses and pumping}

The dissipative dynamics of the emitter requires a bit more care because of the intrinsic nonlinearity of a two-level system. 

We take each emitter to be independently coupled to its own loss bath with a Hamiltonian of the form
\begin{equation}
H_\gamma = \sum_{q}\left( \gamma_q^* S^+A_q + \gamma_q A^{\dagger}_q S^- \right).
\end{equation} 
Here, $q$ indicates the modes of the bath, $\gamma_q$ are the coupling constants, and $A_q$ are the bath operators, assumed to have bosonic nature and an initially very low temperature. 
Performing a Markov approximation, the quantum Langevin equations for the spin-like operators of the emitter read 
\begin{equation}
\begin{cases}
\left.\frac{dS^z}{dt}\right|_\gamma   =   - \gamma\left(\frac{1}{2} + S^z\right) + G^z_\gamma\\
\left.\frac{dS^+}{dt}\right|_\gamma  =  \left(i\nu - \frac{\gamma}{2} \right)S^+ + G^+_\gamma.
\end{cases}
\end{equation}
The deterministic part of these equations shows that each emitter tends to decay towards its lower state independently of its neighbors. Differently from what happened to the cavity mode in \eqref{F1} and \eqref{F2}, the noise operators $G^+_\gamma$ and $G^z_\gamma$ now depend on the initial state of the bath $A_q(t_0)$ as well as on the instantaneous spin operators:
\begin{gather}
\begin{split}
G^z_\gamma(t)  = & -i\sum_q\left[\gamma_q^*\mathrm{e}^{-i\omega_q(t-t_0)}S^+(t)A_q(t_0)\right. + \\
                          &- \left. \gamma_q\mathrm{e}^{i\omega_q(t-t_0)}A_k^{\dagger}(t_0)S^-(t)\right],
\end{split}\\
G^+_\gamma(t) = -2i\sum_k\gamma_k\mathrm{e}^{i\omega_k(t-t_0)}A^{\dagger}_k(t_0)S^z(t).
\end{gather} 
Under the same conditions assumed for the cavity operators, the quantum noises on the different emitters are uncorrelated and have a delta-like temporal correlation,
\begin{equation}
\langle  G_{\gamma,i}^\alpha(t)G_{\gamma,j}^{\alpha'}(t') \rangle = 2D^{\alpha \alpha'}_\gamma(t)\delta(t-t')\,\delta_{ij}. 
\end{equation}
Among the many $\alpha,\alpha' = +,\, -,\, z$ terms, the only non-zero diffusion coefficients are:
\begin{gather}
D^{-+}_\gamma = \frac{\gamma}{2}, \qquad D^{-z}_\gamma = \frac{\gamma}{2}\langle S^-\rangle, \\
\qquad D^{z+}_\gamma = \frac{\gamma}{2}\langle S^+\rangle ,\qquad D^{zz}_\gamma = \frac{\gamma}{2}\left(\frac{1}{2}+\langle S^z\rangle\right).
\end{gather}
The dependence of the diffusion coefficients on the spin operator averages stems from the intrinsic optical nonlinearity of two-level emitter and makes calculations much harder.

The incoherent external pumping of the system is modelled by coupling each emitter with a bath of inverted oscillators as typically done in laser theory~\cite{GARD2000}. This leads to quantum Langevin equations of the form
\begin{equation}
\left\{\begin{array}{lll}
 \left.\frac{dS^z}{dt}\right|_d   =  d\left(\frac{1}{2} - S^z\right) + G^z_d,\\
 \left.\frac{dS^+}{dt}\right|_d  =  \left(i\nu - \frac{d}{2} \right)S^+ + G^+_d
\end{array}
\right.
\end{equation}
Again, the noise operators $G_d^\alpha$ depend on the spin operators and satisfy delta-like correlation functions in time. The only non-zero diffusion coefficients are now:
\begin{gather}
D^{+-}_d = \frac{d}{2}, \qquad D^{+z}_d = -\frac{d}{2}\langle S^+\rangle,\\
\qquad D^{z-}_d = -\frac{d}{2}\langle S^-\rangle, \qquad D^{zz}_d = \frac{d}{2}\left(\frac{1}{2}-\langle S^z\rangle\right).
\end{gather}   
Combining the two loss and pumping contributions to the emitter dissipative dynamics, one finally obtains
\begin{equation}\label{TLA_eq}
\left\{
\begin{array}{lll}
\left.\frac{dS^z}{dt}\right|_{\gamma+d}  & = & \Gamma\left(\frac{\mathcal{D}}{2} - S^z\right) + G^z,\\
\left.\frac{dS^+}{dt}\right|_{\gamma+d}  & = & \left(i\nu - \frac{\Gamma}{2} \right)S^+ + G^+, 
\end{array}
\right.
\end{equation}
where $\Gamma = d + \gamma$ and $G^\alpha(t) = G^\alpha_{\gamma}(t) + G^\alpha_{d}(t) $. The stationary value of the average inversion operator $S^z$ in the absence of any cavity field  can be called \emph{unsaturated population inversion} and depends only on the ratio between damping rates $x = {d}/{\gamma} $,
\begin{equation}
\mathcal{D} = \frac{d - \gamma }{d + \gamma}.
\end{equation}
In the $\alpha,\alpha' = +,\, -,\, z$ basis, the diffusion matrix $D^{\alpha\alpha'}$ of the total external noise operators $G^\alpha$  is given by:
\begin{equation}\label{diff_coeff}
\left(
\begin{array}{ccc}
  0 & \frac{\gamma}{2} &\frac{\gamma}{2}\langle S^+\rangle\\
\frac{d}{2} & 0 &  -\frac{d}{2}\langle S^-\rangle\\
  -\frac{d}{2}\langle S^+\rangle &  \frac{\gamma}{2}\langle S^-\rangle&  \frac{\Gamma}{2}\left(\frac{1}{2} - \mathcal{D}\langle S^z\rangle \right)
\end{array} 
\right).
\end{equation}

\subsection{The quantum Langevin equations}

Putting all terms together, we obtain the final quantum Langevin equations for the $i$-th emitter and the $\kk$ cavity mode {operators},
\begin{gather}
\begin{split}
\frac{dS^z_i}{dt}    =  \Gamma\left(\frac{\mathcal{D}}{2} - S^z_i\right)  + &\frac{g}{\sqrt{V}}\sum_{\kk}\left(\ee^{i\kk\cdot\xx_i}\,S_i^+\,\bh_{\kk} +\right.\\
                                             &\left. +\ee^{-i\kk\cdot\xx_i}\,\bhd_{\kk}\,S_i^- \right)+ G^z_i ,
\end{split}\label{eqn1a}\\
\frac{dS^+_i}{dt}         =  \left(i\nu - \frac{\Gamma}{2} \right)S^+_i -\frac{2g}{\sqrt{V}}\sum_{\kk}\ee^{-i\kk\cdot\xx_i}\,\bhd_{\kk}\,S^z_i + G^+_i,        \label{eqn1b}\\
\frac{d\bhd_{\kk}}{dt}      =  \left( i\omega_{\kk} - \frac{\kappa}{2} \right) \bhd_{\kk} - \frac{g}{\sqrt{V}}\sum_i\ee^{i\kk\cdot\xx_i}\,S^{+}_i + F_{\kk}^{\dagger}    \label{eqn1c}. 
\end{gather}

{These} equations can be rewritten in {{\em real space}} in terms of field and spin-density operators. Assuming the emitters to be arranged on a regular square lattice with density $n_A$ and to have a fictitious size equal to the lattice cell volume $a=n_A^{-1}$, {these latter can be defined as} 
\begin{equation}
 S^\alpha(\xx) = \sum_i\delta_{a}^{(D)}(\xx-\xx_i)\,S_i^\alpha 
\end{equation}
in terms of delta distributions broadened over a spatial area $a$. Assuming that the bosonic field $\phi(\xx)$ is almost constant over a length $\sim a$ allows us to approximate $\delta_{a}^{(D)}(\xx)$ as a delta function, simplifying the algebra of the spin densities and the form of the quantum Langevin equations.
In this representation, the spin algebra in the cartesian $\alpha_i=x,y,z$ basis has the form 
\begin{equation}
[ S^{\alpha_1}(\xx), S^{\alpha_2}(\xx')] = i\varepsilon_{\alpha_1\alpha_2\alpha_3}  S^{\alpha_3}(\xx) \delta^{(D)}_a(\xx-\xx').
\end{equation}
{Summing up}, the {real space} quantum Langevin equations can be written as
\begin{gather}
\begin{split}
\frac{\partial S^z(\xx)}{\partial t}  =  \Gamma \left[n_A\frac{\mathcal{D}}{2} -   S^z(\xx)\right] &+ g\left[ S^+(\xx)\phi(\xx)\right. + \\
                                                                                                     & \left.+\phi\dg(\xx)  S^-(\xx)\right] + G^z(\xx),
\end{split}\label{eq:space1}\\
\frac{\partial S^+(\xx)}{\partial t}    =  \left[i\nu - \frac{\Gamma}{2}\right] S^+(\xx) -2g\phi\dg(\xx)  S^z(\xx) + G^+(\xx), \label{eq:space2}\\
\frac{\partial\phi\dg(\xx)}{\partial t}     =  \left[ i\omega(i\nabla_\xx) - \frac{\kappa}{2}\right]\phi\dg(\xx) - g S^+(\xx) + F^\dagger(\xx). \label{eq:space3}
\end{gather}
with a spatially local noise correlation
\begin{equation}
\langle G^\alpha(t,\xx) G^{\alpha'}(t',\xx')\rangle = D^{\alpha\alpha'}(\xx)\delta_a^{(D)}(\xx-\xx')\delta(t-t'),
\end{equation} 
with
\begin{equation}\label{diff_coeff_x}
\left(
\begin{array}{ccc}
0 &   \frac{\gamma}{2}\, n_A &  \frac{\gamma}{2}\langle S^+(\xx)\rangle\\
 \frac{d}{2}\,n_A &  0 &   -\frac{d}{2}\langle S^-(\xx)\rangle\\
  -\frac{d}{2}\langle S^+(\xx)\rangle & \frac{\gamma}{2}\langle S^-(\xx)\rangle&  \frac{\Gamma}{2}\left(\frac{n_A}{2} - \mathcal{D}\langle S^z(\xx)\rangle \right)
\end{array} 
\right).
\end{equation}

Another useful representation of the previous equations is in {{\em momentum space}}: defining the Fourier transform of the spin-density as 
\begin{equation}
 S^\alpha_\kk = \int \mathrm{d}^d x\, S^\alpha(\xx)\ee^{-i\kk\cdot\xx}, \qquad  S^\alpha(\xx) = \frac{1}{V}\sum_\kk S_\kk^\alpha \ee^{i\kk\cdot\xx},
\end{equation}
we have the spin commutation relations
\begin{equation}
 [ S^{\alpha_1}_\kk, S^{\alpha_2}_{\kk'}] = i\varepsilon_{{\alpha_1}{\alpha_2}{\alpha_3}} S^{\alpha_3}_{\kk+\kk'},
\end{equation}
 and the quantum Langevin equations
\begin{gather}
\begin{split}
\frac{dS^z_{\kk}}{dt}         =  \Gamma\left(\delta_{\kk,0}N_A\frac{\mathcal{D}}{2} -  S^z_{\kk}\right)  + &\frac{g}{\sqrt{V}}\sum_{\qq}\left( S_{\kk - \qq}^+\bh_{\qq} + \right.\\
                    & \left. +\bhd_{\qq} S_{\kk+\qq}^- \right)+ G^z_{\kk}    ,
\end{split}\label{eqn2a}\\
\frac{dS^+_{\kk}}{dt}         =  \left(i\nu - \frac{\Gamma}{2} \right) S^+_{\kk} -2\frac{g}{\sqrt{V}}\sum_{\qq}\bhd_{\qq} S^z_{\kk+\qq} + G^+_{\kk}      ,  \label{eqn2b}\\
\frac{d\bhd_{\kk}}{dt}      =  \left( i\omega_{\kk} - \frac{\kappa}{2} \right) \bhd_{\kk} - \frac{g}{\sqrt{V}} S^{+}_{-\kk} + F_{\kk}^{\dagger}       \label{eqn2c}. 
\end{gather}
Momentum space noise operators then satisfy 
\begin{gather}
\langle G^\alpha_{\kk}(t)G^{\alpha'}_{\kk'}(t')\rangle = 2D_{\kk+\kk'}^{\alpha\alpha'}\delta(t-t'),\\
\end{gather} 
with
\begin{equation}\label{diff_coeff_k}
\left(
\begin{array}{ccc}
0 &   \frac{\gamma}{2}\, N_A\delta_{\kk,-\kk'} &  \frac{\gamma}{2}\langle S^+_{\kk+\kk'}\rangle\\
 \frac{d}{2}\,N_A\delta_{\kk,-\kk'} &  0 &   -\frac{d}{2}\langle S^-_{\kk+\kk'}\rangle\\
  -\frac{d}{2}\langle S^+_{\kk+\kk'}\rangle & \frac{\gamma}{2}\langle S^-_{\kk+\kk'}\rangle&  \frac{\Gamma}{2}\left(\frac{N_A}{2}\delta_{\kk,-\kk'} - \mathcal{D}\langle S^z_{\kk+\kk'}\rangle \right)
\end{array} 
\right).
\end{equation}
{Before proceeding with our discussion, it is worth pointing out that what we have introduced so far is a minimal quantum model to describe condensation in a spatially extended geometry. Depending on the specific system under investigation, other terms might be needed, for instance dephasing of the emitter under the effect of a sort of collisional broadening, or several species of emitters with different resonance frequencies $\nu_i$ so to account for more complex gain spectra. 

In our formalism, dephasing corresponds to terms of the form
\begin{equation}
\label{eq:MEcoll}
\dot{\rho} = \frac{\Gamma_{\text{coll}}}{2}\left(4S^z\rho S^z - \rho\right)
\end{equation}
in the master equation~\cite{COHE2004}, $\Gamma_\text{coll}$ being the contribution of the dephasing to the dipole relaxation rate. In the quantum Langevin formalism, these processes give additional deterministic terms  
\begin{equation}
\label{QLcoll}
\begin{cases}
\left.\frac{dS^+}{dt}\right|_{\rm coll} = -\Gamma_\text{coll}S^+ + G_{\rm coll}^+, \\
\left.\frac{dS^z}{dt}\right|_{\rm coll} =0,
\end{cases}
\end{equation}
and an additional contribution to the noise:
 \begin{equation}
 \label{QLdiff}
 \begin{cases}
 \langle G_{\rm coll}^+(t)G_{\rm coll}^-(t') \rangle = 2\Gamma_\text{coll}\left(\frac{1}{2} + \langle S^z\rangle\right)\delta(t-t'),\\
  \langle G_{\rm coll}^-(t)G_{\rm coll}^+(t') \rangle = 2\Gamma_\text{coll}\left(\frac{1}{2} - \langle S^z\rangle\right)\delta(t-t').
 \end{cases}
 \end{equation}
We have checked that including such terms does not introduce any qualitatively new feature in the model.}

\section{Mean-field theory}
\label{sec:MF}


{As a first step in our study of non-equilibrium condensation effects, we study the mean-field solution to the quantum Langevin equations.} This amounts to neglecting the quantum noise terms in \eqref{eqn2a}-\eqref{eqn2c} and replacing each operator with its expectation value. This study is the simplest in momentum representation, where the mean-field motion equations for $\beta_\kk^* = \langle \bhd_\kk\rangle$ and $\sigma^\alpha_\kk = \langle S^\alpha_\kk \rangle$ have the form
\begin{gather}
\begin{split}
\dot{\sigma}^z_{\kk}     =  \Gamma\left(\delta_{\kk,0}N_A\right.&\left.\frac{\mathcal{D}}{2} - \sigma^z_{\kk}\right)  + \\
                                        &+\frac{g}{\sqrt{V}}\sum_{\qq}\left(\sigma_{\kk - \qq}^+\beta_{\qq} + \beta_{\qq}^*\sigma_{\kk+\qq}^- \right)  \label{MF1a} ,
\end{split}\\
\dot{\sigma}^+_{\kk}         =  \left(i\nu - \frac{\Gamma}{2} \right)\sigma^+_{\kk} -2\frac{g}{\sqrt{V}}\sum_{\qq}\beta_{\qq}^*\sigma^z_{\kk+\qq}  ,       \label{MF2a}\\
\dot{\beta}_{\kk}^*              =  \left( i\omega_{\kk} - \frac{\kappa}{2} \right) \beta_{\kk}^* - \frac{g}{\sqrt{V}}\sigma^{+}_{-\kk} +\frac{i\lambda}{V}\sum_{\qq\qq'}\beta^*_{\qq+\qq'} \beta^*_{\kk-\qq'} \beta_{\qq}, \label{MF3a}
\end{gather}
very similar to the ones of the semi-classical theory of lasers~\cite{SCUL1997}. 

\subsection{Stationary state: Bose condensation}

While a trivial solution with all $\beta_\kk^*=\sigma_\kk^+=0$ is always present, for some values of the parameters to be specified below, {this solution becomes dynamically unstable and is replaced by} other {\em condensed} solutions with a non vanishing field amplitude. Inspired by experiments, we focus our attention on the case where condensation occurs on the $\kk=0$ state. This corresponds to inserting the ansatz
\begin{equation}\label{ansatz}
\begin{cases}
\beta^*_{\kk}(t) = \delta_{\kk0}\,\sqrt{V}\,\beta_0^*\,\ee^{i\omega t},\\
\sigma^+_{\kk}(t) = \delta_{\kk0}\,V\,\sigma_0^+\ee^{i\omega t}, \\
 \sigma^z_{\kk}(t) = \delta_{\kk0}\,V\,\sigma^z_0,
\end{cases}
\end{equation}
into the mean-field equations, with the amplitudes $\beta_0^*$ and $\sigma_0^+$, the population inversion $\sigma^z_0$ and the frequency $\omega$ to be determined in a self consistent way. 

In the $\lambda=0$ case where direct photon-photon interactions vanish, a direct analytical solution of the mean-field equations gives
\begin{equation}\label{freq}
 \omega = \frac{\frac{\nu}{\Gamma} + \frac{\omega_{0}}{\kappa}}{\frac{1}{\Gamma} + \frac{1}{\kappa} } = \omega_0 + \frac{\kappa}{2}\delta, 
\end{equation}
where $\delta = 2(\nu - \omega_0)/(\Gamma + \kappa)$ is the dimensionless detuning: the frequency $\omega$ is therefore equal to an average of the bare field and dipole frequencies, weighted with their bare lifetimes.
Analogously, we find for the field and emitter observables,
\begin{eqnarray}\label{mfvalues}
|\beta_0|^2          & = &   \frac{\Gamma}{\kappa}\left[ n_A\frac{\mathcal{D}}{2} - \frac{\Gamma\kappa}{8g^2}\left(1 + \delta^2 \right) \right], \label{field}\\
\sigma^z_0      & = &   \frac{\Gamma\kappa}{8g^2}\left(1 + \delta^2\right),            \label{inversion}\\
\sigma^+_0      & = &   -\frac{\kappa}{2g}\left(1 + i\delta\right)\beta_0^*. \label{dipole}
\end{eqnarray}
The condensation threshold is clearly visible in these results: for {$\mathcal{D}/\Gamma<\kappa(1+\delta^2)/4g^2n_A$}, the right-hand side of \eqref{field} is negative, so only the trivial $\beta_0$ solution is possible. For {$\mathcal{D}/\Gamma>\kappa(1+\delta^2)/4g^2n_A$}, a condensed solution appears with a finite field intensity \eqref{field} and a corresponding emitter dipole moment proportional to \eqref{dipole}. {Remind that both $\mathcal{D}=(d-\gamma)/(d+\gamma)$ and $\Gamma=d+\gamma=\gamma(1+x)$ are here function of the pumping rate.} 

For finite values of $\lambda$, a similar derivation can be carried out. For the frequency, it gives
\begin{equation}
\label{shift}
 \omega = \frac{\frac{\nu}{\Gamma} + \frac{1}{\kappa}\left(\omega_0 + \lambda|\beta_0|^2\right)}{\frac{1}{\Gamma} + \frac{1}{\kappa}} = \omega_0 + \lambda|\beta_0|^2 + \frac{\kappa}{2}\delta_\lambda,
\end{equation}
where the dimensionless detuning $ \delta_\lambda = 2\left(\nu-\omega_0 - \lambda|\beta_0|^2\right)/\left(\Gamma+\kappa\right)$ now involves also the nonlinear frequency shift of the field mode. For the field and the emitter observables, it gives:
\begin{eqnarray}
|\beta_0|^2    & = &   \frac{\Gamma}{\kappa}\left[ n_A\frac{\mathcal{D}}{2} - \frac{\Gamma\kappa}{8g^2}\left(1 + \delta_\lambda^2 \right) \right] \label{eq:MFint1}\\
\sigma^z_0      & = &   \frac{\Gamma\kappa}{8g^2}\left(1 + \delta_\lambda^2\right)            \label{eq:MFint2}\\
\sigma^+_0      & = &   -\frac{\kappa}{2g}\left(1 + i\delta_\lambda\right)\beta_0^*. \label{eq:MFint3}
\end{eqnarray}

\subsection{Physical discussion}
 \begin{figure*}[htbp]
 \begin{center}
 \includegraphics[width=18 cm]{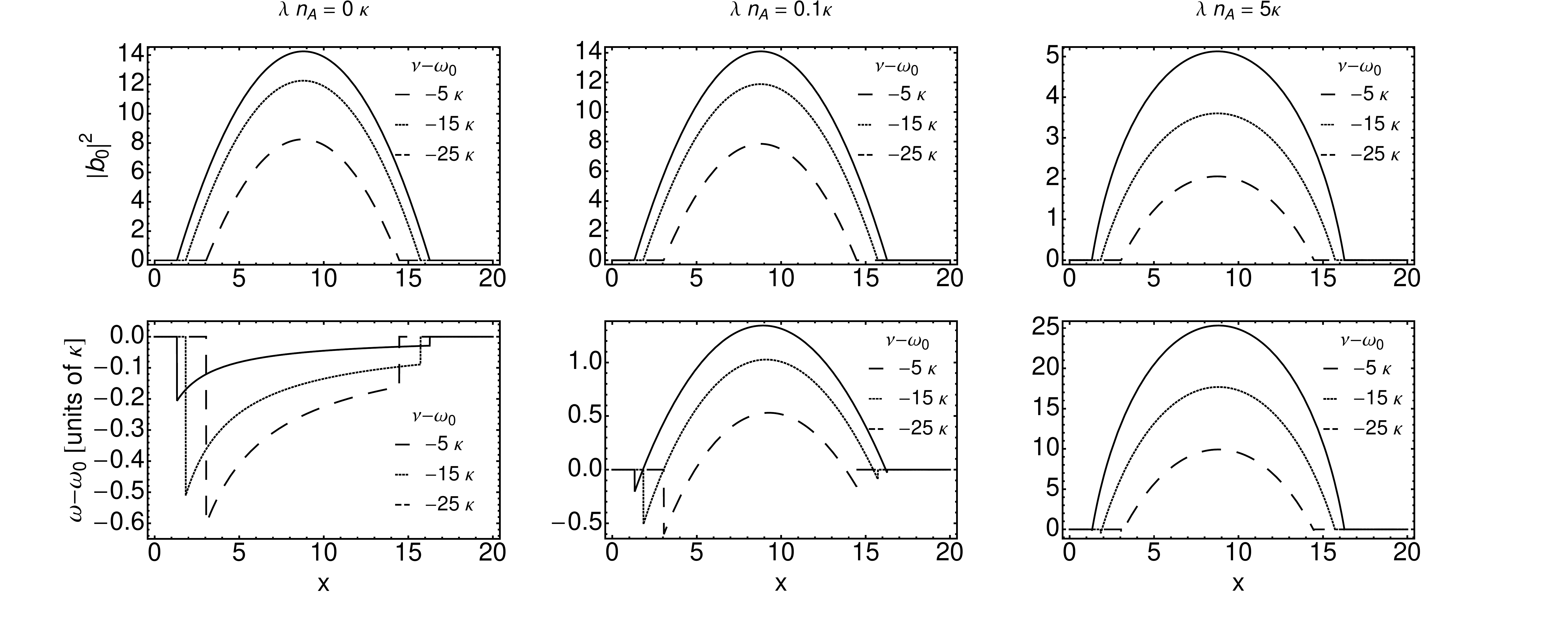}
 \caption{Intensity of the field (upper panels) and {oscillation} frequency of the condensate (lower panels) as a function of the pumping parameter $x = d/\gamma$. Both quantities are shown for different values of self-interaction $\lambda$ and natural detuning $\nu-\omega_0$. {In all panels,} $\gamma = 10 \kappa$ and $g\sqrt{n_A} = 7 \kappa$.} 
 \label{fig:1}
 \end{center}
 \end{figure*}

The most remarkable feature of the mean-field equations is the spontaneous symmetry breaking phenomenon at the condensation threshold. The mean-field equations \eqref{MF1a}-\eqref{MF3a} are symmetric under the U(1) transformation $ (\beta^*_{\kk},\sigma^+_{\kk}) \rightarrow (\ee^{i\varphi}\beta^*_{\kk},\ee^{i\varphi}\sigma^+_{\kk})$ with arbitrary global phase $\varphi$. While for all values of the parameters there is a trivial $\beta_0=\sigma^+_0=0$ solution which fulfils this symmetry, any non-trivial  solution has to choose a specific phase for $\beta_0$ and $\sigma_0^+$, only their modulus being fixed by \eqref{field} or \eqref{eq:MFint1}: as a result, the U(1) symmetry is {\em spontaneously broken}. In actual experiments, this phase is randomly chosen. Note that since the symmetry transformation does not involve $\sigma^z_{0}$, its mean-field value can always be non-zero.

The behaviour of the field intensity $|\beta_0|^2$ and of the oscillation frequency $\omega$ is plotted in Fig.\ref{fig:1} as a function of the pumping strength $x=d/\gamma$ for different (negative) values of the natural field-emitter detuning $\nu-\omega_0<0$ (different curves) and different values of the (positive) nonlinear coupling $\lambda>0$ (different panels). 
In all cases, two thresholds are well visible: the lower one corresponds to the standard switch-on of laser operation for sufficiently large pump strength. The upper one is a consequence of our specific model and is due to the fact that the gain offered by the emitters is suppressed when the effective emitter linewidth $\Gamma=d+\gamma=\gamma(1+x)$ appearing in \eqref{eqn1b} is very much broadened by the pumping term $d$. As usual, whenever a non-trivial  $\beta_0\neq 0$ condensate solution is available, the trivial solution becomes dynamically unstable. For all cases shown in this figure, the order parameter $\beta_0$ grows continuously from zero, so the condensation resembles a second-order phase transition.

The behavior of the oscillation frequency shown in the lower panels of Fig.\ref{fig:1} is determined by a complex interplay of the bare frequencies of the cavity and of the emitter, weighted by their respective linewidths and shifted by the nonlinear interaction energy $\lambda$ according to \eqref{shift}.

The situation for positive detuning $\nu-\omega_0>0$ is more complicate and a complete analysis of the rich phenomenology goes beyond the scope of this work. {Not only the order parameter as a function of pumping strength can be discontinuous~\cite{Wouters2007} and bistable, but also the spatial shape of the condensate can develop a complicate structure. As the gain is maximum on a $\kk$-space ring of modes at a finite $k$, the choice of the specific combination of modes is determined by complex mechanisms involving the interplay of pumping and dissipation, but also the geometrical details of the system beyond the idealised spatially homogeneous approximation. This complex physics is typical of non-equilibrium systems where no minimal free-energy criterion is available to determine the steady state of the system and is closely related to pattern formation in nonlinear dynamical systems~\cite{Cross}. First experimental evidence of condensation in spatially non-trivial modes was reported in~\cite{Richard:PRL2005} and discussed in~\cite{WOUT2008}. More complicate spatial features were investigated in~\cite{Keeling,BAUM}.}

\section{Quantum fluctuations}
\label{sec:Fluctuations}

\subsection{{Linearised theory of small fluctuations}}

The mean-field steady-state solution obtained in the previous Section is the starting point for a linearised theory of fluctuations. 
In the spirit of Bogoliubov and the spin wave approximations, we can linearise Eqs.\eqref{eqn2a}-\eqref{eqn2c} around the steady-state by performing the operator replacement:
\begin{equation}
\label{linear}
\begin{cases}
 \bhd_\kk = \left(\delta_{\kk0}\,\sqrt{V}\,\beta_0^* + \delta \bhd_\kk\right)\ee^{i\omega t}, \\
  S^+_\kk = \left(\delta_{\kk0}\,V\,\sigma^+_0 + \delta S^+_\kk\right)\ee^{i\omega t}, \\
   S^z_\kk = \delta_{\kk0}\,V\,\sigma^z_0 + \delta S^z_\kk:
 \end{cases}
\end{equation}
$\beta_0^*$, $\sigma^+_0$ and $\sigma^z_0$ are here the mean-field steady-states as defined in \eqref{eq:MFint1}-\eqref{eq:MFint3} with a frequency $\omega$ determined by \eqref{shift}. Fluctuations around the mean-field are described by the $\delta \bhd_\kk$, $\delta S^+_\kk$ and $\delta S^z_\kk$ operators which inherit the commutation rules from the original $\bhd_\kk$, $S^+_\kk$ and $S^z_\kk$ operators.

Substituting the previous expressions into the motion equations \eqref{eqn2a}-\eqref{eqn2c} and neglecting terms of second or higher order in the fluctuation operators, we obtain a set of coupled linear equations
\begin{equation}\label{stoceq}
 \frac{d\mathbf{v}_\kk}{dt} = \mathbb{A}_\kk\mathbf{v}_\kk + \widetilde{\mathbf{F}}_\kk,
\end{equation}
for the (rescaled) fluctuation vector
\begin{equation}
\mathbf{v}_\kk^t = ( \delta \widetilde{\bh}^\dagger_{-\kk},  \delta \widetilde{\bh}_{\kk},  \delta S_\kk^+, \delta S_\kk^- , \delta S_\kk^z ),
\end{equation}
with a quantum noise vector
\begin{equation}
\widetilde{\mathbf{F}}_\kk^t = (\widetilde{F}_{-\kk}^{\dagger},\widetilde{F}_\kk, \widetilde{G}_\kk^+, \widetilde{G}_\kk^-, \widetilde{G}_\kk^z).
\end{equation}
For notational convenience, we have used the rescaled quantities $\delta \widetilde{\bh}^\dagger_\kk = \sqrt{V}\delta \bhd_\kk$ with rescaled noise terms $\widetilde{F}^\dagger_\kk = \sqrt{V} \ee^{-i\omega t}F_\kk^\dagger$ and $\widetilde{G}^+_\kk = \ee^{-i\omega t}G_\kk^+$ and $\widetilde{G}^z_\kk = G_\kk^z$. The equations for the Hermitian conjugate quantities $\dsm$ and $\delta\bh_\kk$ follow straightforwardly from $\delta S_{-\kk}^- = (\dsp)^\dagger$ and $\delta\bh_\kk = (\delta \bh^\dagger_{-\kk})^\dagger$.

Defining the shorthands $z_\lambda = 1 + i\delta_\lambda$  and $\epsilon_\kk =  k^2/2m$, the Bogoliubov matrix $\mathbb{A}_\kk$ is equal to
\begin{widetext}
\begin{equation}
\mathbb{A}_\kk = \left(
\begin{array}{ccccc} 
-\frac{\kappa}{2}z_\lambda + i \epsilon_\kk  + i\lambda|\beta_0|^2 & i\lambda (\beta_0^*)^2                                                 & -g               & 0                                     & 0                  \\
-i\lambda \beta_0^2                                                  & -\frac{\kappa}{2}z_\lambda^* - i \epsilon_\kk - i\lambda|\beta_0|^2  & 0                                       & -g             & 0                  \\
-2g\sigma_0^z                                           & 0                                                                & -\frac{\Gamma}{2}z_\lambda^*            & 0                                     & -2g \beta_0^*        \\
0 		   		                               & -2g\sigma_0^z                                             & 0                                       & -\frac{\Gamma}{2}z_\lambda            & -2g \beta_0          \\
g\sigma_0^-	                                       & g\sigma_0^+		                                  & g \beta_0                                 & g \beta_0^*                             & -\Gamma      
\end{array}   
\right).
\end{equation}
\end{widetext}

Evaluation of the noise correlation matrix requires a bit more care as the emitter noise depends on the emitter operators themselves. 

Inserting into \eqref{diff_coeff_k} the {steady-state value of the emitter operators}, we have that:  
\begin{equation}
 \langle\widetilde{G}_\kk^\alpha\widetilde{G}_{\kk'}^{\alpha'} \rangle = 2D_{\kk+\kk'}^{\alpha\alpha'}\,\delta(t-t') \delta_{\kk+\kk',0} \propto N_A:
 \label{noiseG}
\end{equation}
as in this equation the emitter noise terms $G_\kk^\alpha \propto \sqrt{N_A}$ are of the same order as the other terms in the linearised equations, it is legitimate to replace the spin operators in the diffusion coefficients with their mean field values. 
Note that the $\delta_{\kk+\kk',0}$ coefficient in \eqref{noiseG} is a consequence of the assumed ordered arrangement of the emitters: Had we considered a disordered configuration, the zero value for $\kk+\kk'\neq 0$ would be replaced by something proportional to $\sqrt{N_A}$, still negligible with respect to the value proportional to $N_A$ of the $\kk+\kk'=0$ term.

The correlation matrix of $\widetilde{\mathbf{F}}_\kk$ is
\begin{equation}
\langle \widetilde{\mathbf{F}}_\kk(t) \widetilde{\mathbf{F}}_{\kk'}^\dagger(t') \rangle = \mathbb{D}\delta(t-t')\delta_{\kk,\kk'}
\end{equation}
with
\begin{equation}\label{newdiff}
 \mathbb{D} = V\left(
\begin{array}{ccccc}
0                               & 0                             & 0                  & 0                  & 0                                                  \\
0                               & \kappa                        & 0                  & 0                  & 0                                                  \\
0                               & 0                             & d\, n_A              & 0                  & -d \sigma_0^+                                           \\
0 				& 0                             & 0                  &\gamma \, n_A         & \gamma \sigma_0^-                                      \\
0				& 0      			& -d\sigma_0^-           & \gamma \sigma_0^+      & \Gamma\left(\frac{n_A}{2} - \mathcal{D}\sigma_0^z \right) \\     
\end{array}
\right).
\end{equation}

As a final remark on the linearisation procedure, let us emphasize how our approximations are controlled by the total number of atoms $N_A$. Assume the scaling
\begin{equation}
S_{\kk=0}^\alpha \sim N_A, \quad \bh_{\kk=0} \sim \sqrt{N_A}, \quad D^{\alpha\alpha'}_{\kk=0} \sim N_A,
\end{equation}    
and
\begin{equation}
S_{\kk\neq0}^\alpha \sim \sqrt{N_A}, \quad \bh_{\kk\neq0} \sim 1, \quad D^{\alpha\alpha'}_{\kk\neq0} \sim \sqrt{N_A}, 
\end{equation}    
together with $g\sim 1/\sqrt{n_A}$ the dependence on $N_A$ of each term in Eqs. \eqref{eqn2a}-\eqref{eqn2c} can be made explicit. Then, in the thermodynamical limit $N_A\to+\infty$, retaining the leading order in $N_A$ from such equations is equivalent to the perform mean-field approximation of Sec. \ref{sec:MF}. If the next-to-leading order is also retained, the linearised Bogoliubov theory is recovered. 

{In analogy with the systematic expansion of equilibrium Bogoliubov theory in powers of the dilution parameter~\cite{castindum}, we can make use of these considerations to define a systematic {\em mean-field} limit for our non-equilibrium system. To this purpose, it is useful to consider the real-space form of the quantum Langevin equations (\ref{eq:space1}-\ref{eq:space3}). If we let the atomic density and the photon density $|\phi(\xx)|^2\sim S^\alpha(\xx)\sim n_A\to \infty$ at constant $g\sqrt{n_A}\sim g|\phi(\xx)|$ and $\lambda |\phi(\xx)|^2$, the mean-field equations are not affected [in particular, their steady-states (\ref{eq:MFint1}-\ref{eq:MFint3})], while the relative importance of the noise terms in the quantum Langevin and of the commutators tends to zero. As a result, the relative magnitude of quantum fluctuation expectation values vs. mean-field terms scale as $1/n_A$ in the mean-field limit.}

\subsection{The collective Bogoliubov modes}

\begin{figure}[htbp]
   \includegraphics[width=8.5cm]{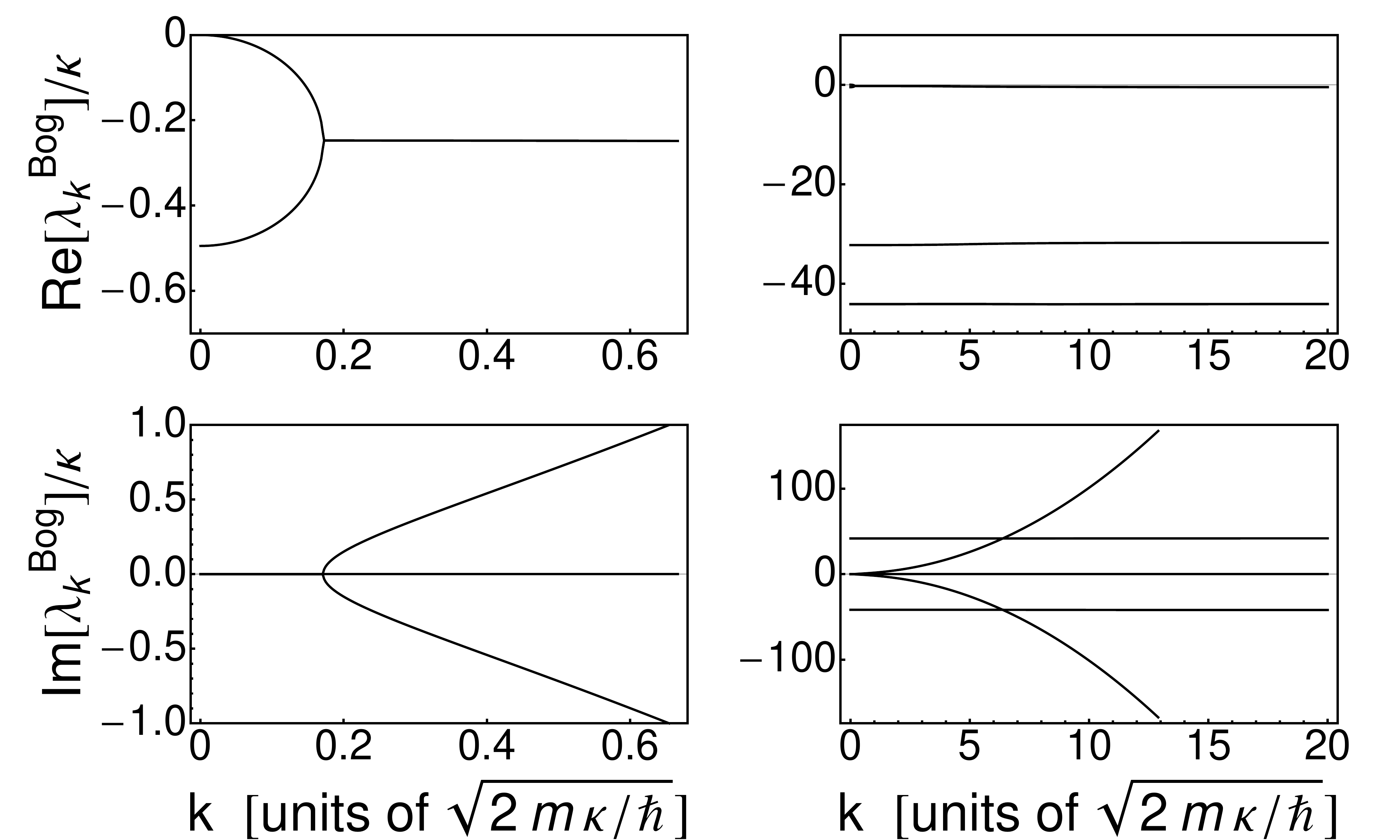}
   \caption{ {Dispersion $\lambda_k^{\mathrm{Bog}}$ of the collective modes as predicted by the eigenvalues of the Bogoliubov matrix $\mathbb{A}_\kk$ in the interacting case with $\lambda n_A = 0.1\kappa$ and $\nu-\omega_0 = -10 \kappa$. Left panels show magnified views of the low-$\kk$ region of right panels. System parameters: $\gamma = 100 \kappa$, $g\sqrt{n_A} = 25$ and $x = 5$.}}
    \label{fig:2}
 \end{figure}

\begin{figure}[htbp]
   \includegraphics[width=8.5cm]{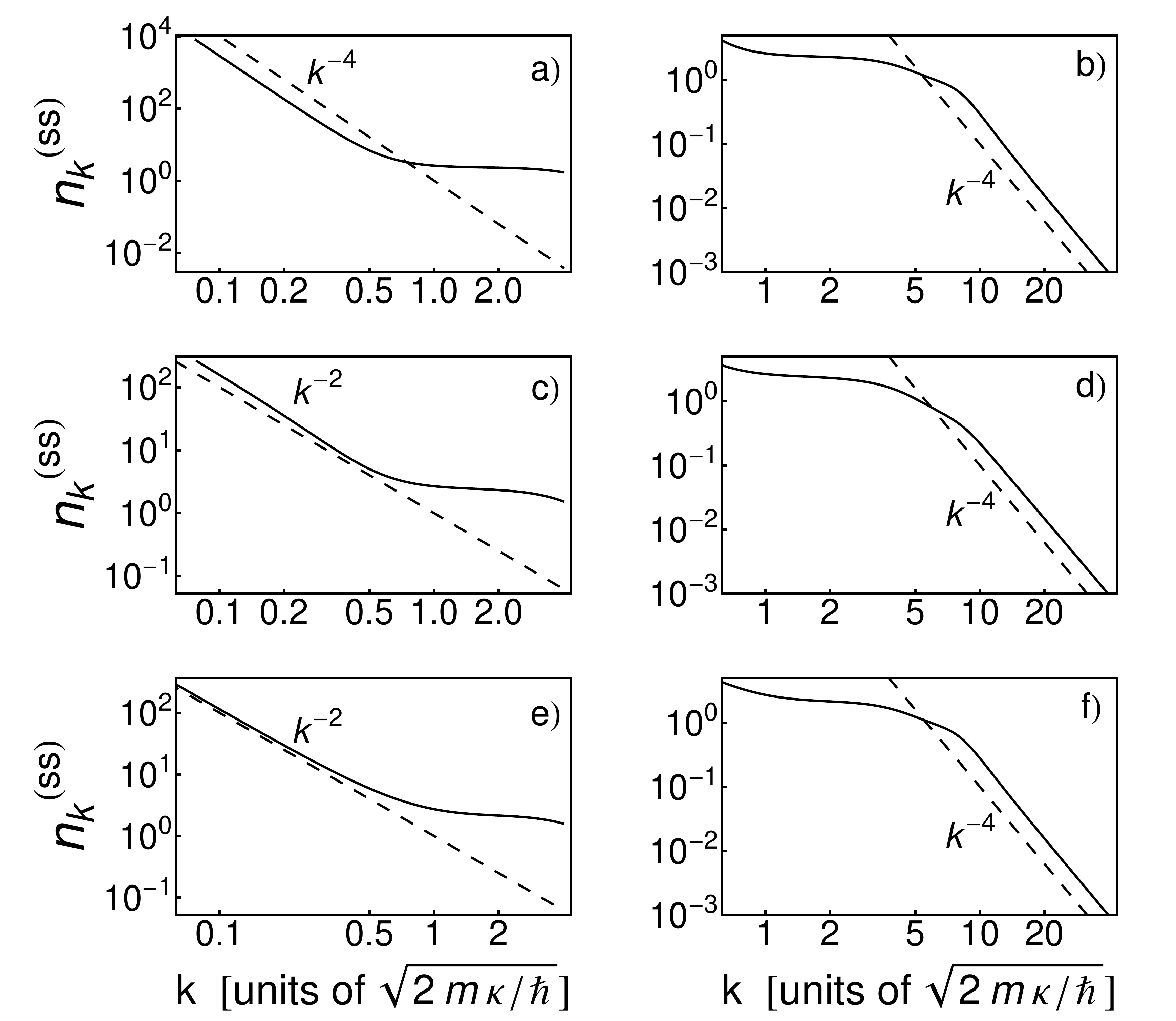}
   \caption{{Steady state momentum distribution. Left panels show magnified views of the low-$\kk$ region of right panels. (a,b) non-interacting case $\lambda n_A = \nu-\omega_0 =0$. (c,d) $\lambda n_A =0$, $\nu-\omega_0 = -10\kappa$. (e,f) $\lambda n_A =0.1\kappa$, $\nu-\omega_0 = 0$. System parameters: $\gamma = 100 \kappa$, $g\sqrt{n_A} = 25$ and $x = 5$.}}
    \label{fig:3}
 \end{figure}

A first step to physically understand the consequences of fluctuations is to study the dispersion of the eigenvalues $\lambda^{\rm Bog}_\kk$ of $\mathbb{A}_\kk$ as a function of $k$, which gives the generalised Bogoliubov dispersion of excitations on top of the non-equilibrium condensate. 

{An example of dispersion is shown in Fig.\ref{fig:2}: the upper panels show the real part of the dispersion {$\textrm{Re}[\lambda^{\bf Bog}_k]$ (describing the damping/growth rate of the mode)} and the lower panels show the imaginary part {$\textrm{Im}[\lambda^{\rm Bog}_k]$ (describing the oscillation frequency of the mode)}. The left column give magnified views of the same dispersion shown on the right column.}
%
%
%
%
%

As expected there is a Goldstone mode corresponding to the spontaneously broken U(1) symmetry, whose frequency tends to $0$ in both real and imaginary parts as $k\to 0$. As typical in non-equilibrium systems~\cite{Cross}, this mode is however diffusive rather than sonic, that is {$\textrm{Im}[\lambda_k^{\textrm{Bog}}]=0$} for a finite range around $k=0$ and the {real} part starts from zero as {$\textrm{Re}[\lambda_k^{\textrm{Bog}}]\simeq-\zeta k^2$}. 

At higher momenta, the diffusive Goldstone mode transform itself into a single-particle cavity photon mode with a parabolic dispersion. Between the two regimes, for $\lambda>0$ or a finite cavity-emitter detuning $\delta$, there is a sonic-like dispersion of the {$\textrm{Im}[\lambda_k^{\textrm{Bog}}] \approx c_s |k|$} form (see Figs.\ref{fig:2}): for $\lambda>0$, this is a standard feature of the Bogoliubov dispersion of interacting photons/polaritons~\cite{RMP2013}. For a finite $\delta$, it follows from the intensity-dependence of the refractive index of detuned two-level systems~\cite{COHE2004}. A connection with the Gross-Pitaevskii formulation of \cite{WOUT2007} will be given at the end of Sec.\ref{sec:SGPE}.

In the larger view displayed on the right column, in addition to the Goldstone mode we see two other, almost dispersionless excitation modes. As their origin is mostly due to emitter degrees of freedom, they could not be captured by the Gross-Pitaevskii approach of~\cite{WOUT2007}. Their splitting is related to the Rabi frequency of the optical dressing of the atoms due to the coherent field in the cavity corresponding to the condensate and they are visible in the emitter emission spectrum as the external sidebands of the so-called Mollow triplet of resonance fluorescence~\cite{COHE2004}.

\begin{figure}[htbp][h]
 \begin{center}
 \includegraphics[width=8.5 cm]{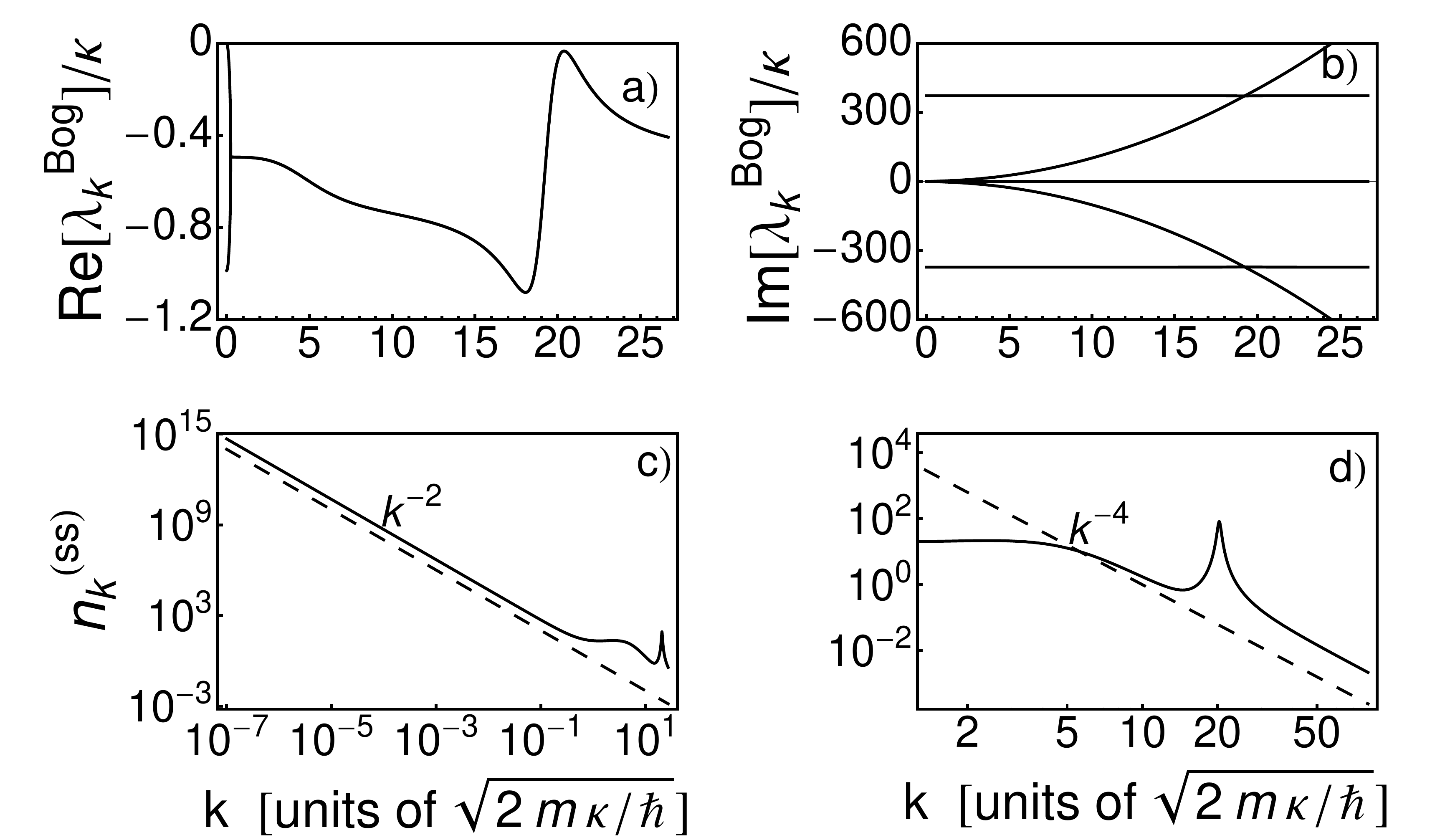}
 \caption{{Imaginary (a) and real (b) part of the dispersion of the collective modes and steady state momentum distribution (c)-(d) in the vicinity of the Mollow instability onset. (c) shows a magnified view of the low-$\kk$ region of (d). System parameters: $\gamma = 10 \kappa$, $g\sqrt{n_A} = 42 \kappa$, $x = 5$, $\lambda n_A = 0.1\kappa$, $\nu-\omega_0=0 $.}
}
\label{fig:6}
 \end{center}
 \end{figure}

The effect of these additional modes is more evident in Fig.\ref{fig:6}, where the chosen parameters are close to a secondary instability. The finite instability wavevector is located at the point where the cavity field dispersion crosses the ones of the dispersionless modes: in this neighborhood, the {real} part of the dispersion {$\textrm{Re}[\lambda_\kk^{\rm Bog}]$} approaches $0$ from below. Should {$\textrm{Re}[\lambda_\kk^{\rm Bog}]$} go above $0$, our ansatz with a uniform condensate localised in the $\kk=0$ mode would no longer be valid and more complicate condensate shapes with spatial modulation should be considered{~\cite{Keeling,Sarchi:2010PRB,supersolid2,supersolid1}, analogous to secondary instabilities in pattern formation theory~\cite{Cross}}. Physically, this {\em Mollow instability} can be easily interpreted in terms of the well-known optical gain offered by a two-level emitter driven by a strong coherent beam and probed by a weak probe beam detuned by approximately the Rabi frequency of the dressing~\cite{COHE2004}.

\subsection{Momentum distribution}

From the quantum Langevin equation \eqref{stoceq}, it is straightforward to extract predictions for one-time physical observables.
As a most remarkable example, here we shall concentrate our attention on the steady-state momentum distribution of the cavity field,
\begin{equation}
n^s_\kk=\langle \bhd_\kk \,\bh_\kk\rangle=\langle \delta \bhd_\kk \,\delta \bh_\kk\rangle.
\label{nskk}
\end{equation}
On one hand, in contrast to the mean-field approximation where the cavity field is concentrated in the $\kk=0$ mode, this observable {is a sensitive probe of} fluctuations. On the other hand, it is an experimentally accessible quantity, easily measured from the far-field angular distribution of emitted light. By Fourier transform, it is directly related to the two-points, one-time coherence function of the cavity field, a quantity which is of widespread use in experiments~\cite{KASP2006,BAAS2006,ROUM2012,SPAN2012}.

Grouping in the $\mathbb{V}_\kk  = \langle \mathbf{v}^s_\kk{\mathbf{v}^s}^\dagger_\kk\rangle$ variance matrix the steady-state variances of all operator pairs, from a straightforward integration of the quantum Langevin equations~\cite{GARD2004}, we obtain a Lyapunov equation:
\begin{equation}\label{eqvar}
 \mathbb{A}_\kk\mathbb{V}_\kk + \mathbb{V}_\kk\mathbb{A}_\kk^\dagger = -\mathbb{D}
\end{equation}
from which standard linear algebra methods allow to extract the variance matrix $\mathbb{V}_\kk$.

While no simple analytical form is available for $n^s_\kk$, plots of its behaviour are given in {the bottom panels of} Fig. \ref{fig:3} for several most relevant cases. For small $k$, the momentum distribution follows the same  $1/k^2$ behaviour as equilibrium systems provided photons are effectively interacting, that is either $\lambda>0$ or $\delta\neq 0$. In the $\lambda=\delta=0$ case, the situation is more complicate and the distribution appears to diverge as $1/k^4$. Both these results are in agreement with the predictions of the stochastic Gross-Pitaevskii equation in~\cite{ACIC2013}. However, as it was noted in~\cite{Wouters2013}, great care has to be paid when applying the linearised Bogoliubov-like formalism to low-$k$ modes in non-equilibrium, as the effects beyond linearisation can play a dominant role.

At large $k$, the momentum distribution always decays to zero as $1/k^4$. The large-$k$ decay qualitatively recovers the prediction we {guessed} in~\cite{ACIC2013} from a phenomenological stochastic Gross-Pitaevskii equation with a frequency-dependent pumping. The specific $1/k^4$ law is a consequence of our choice of monochromatic emitters, whose amplification spectrum decays as $1/(\omega-\nu)^2$: other choices of the emitter distribution would lead to correspondingly different high-momentum tails of $n_\kk$. The {\em ab initio} confirmation of this large-$k$ decay of $n_\kk$ is one of the main results of this article, as it shows that thermal-like momentum distributions can be found also in models where the quasi-particles are not interacting at all and therefore can not get thermalised by collisional processes. A similar feature was experimentally observed in~\cite{BAJO2007} using a VCSEL device in the weak coupling regime where photons are practically non-interacting.

The intermediate-$k$ region shows a quite structureless plateau connecting the low-$k$ and high-$k$ regimes. The most interesting feature in this window is the peak that appears at the crossing point of the Goldstone mode and the dispersionless branch when the Mollow instability is approached, see Fig.\ref{fig:7}. As usual, the peak height diverges at the onset of the instability.

\subsection{Photo-luminescence spectrum}
 \begin{figure*}[htbp]
 \subfigure[]{\includegraphics[width=7cm]{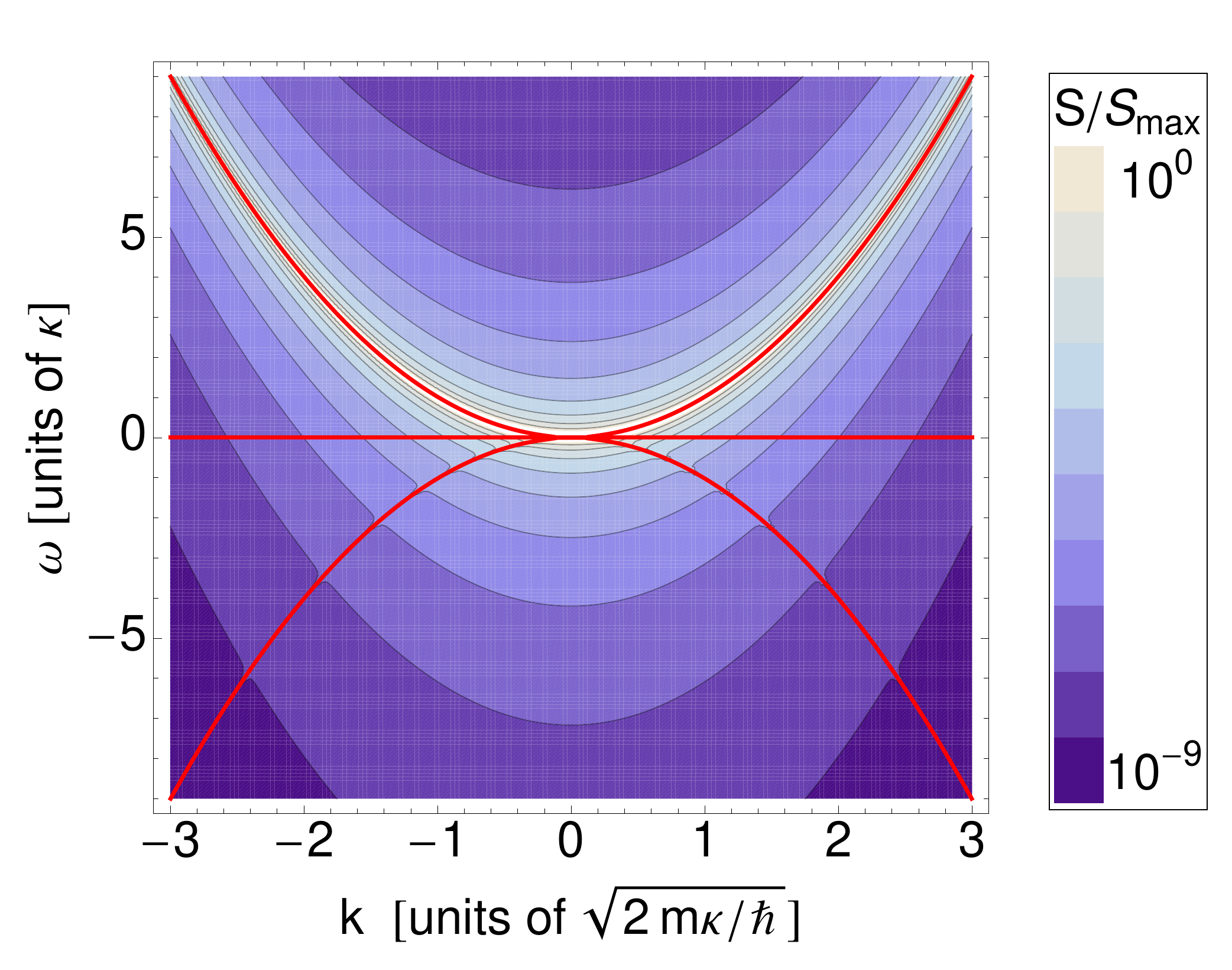}}
  \hspace{0.5cm}
 \subfigure[]{\includegraphics[width=7cm]{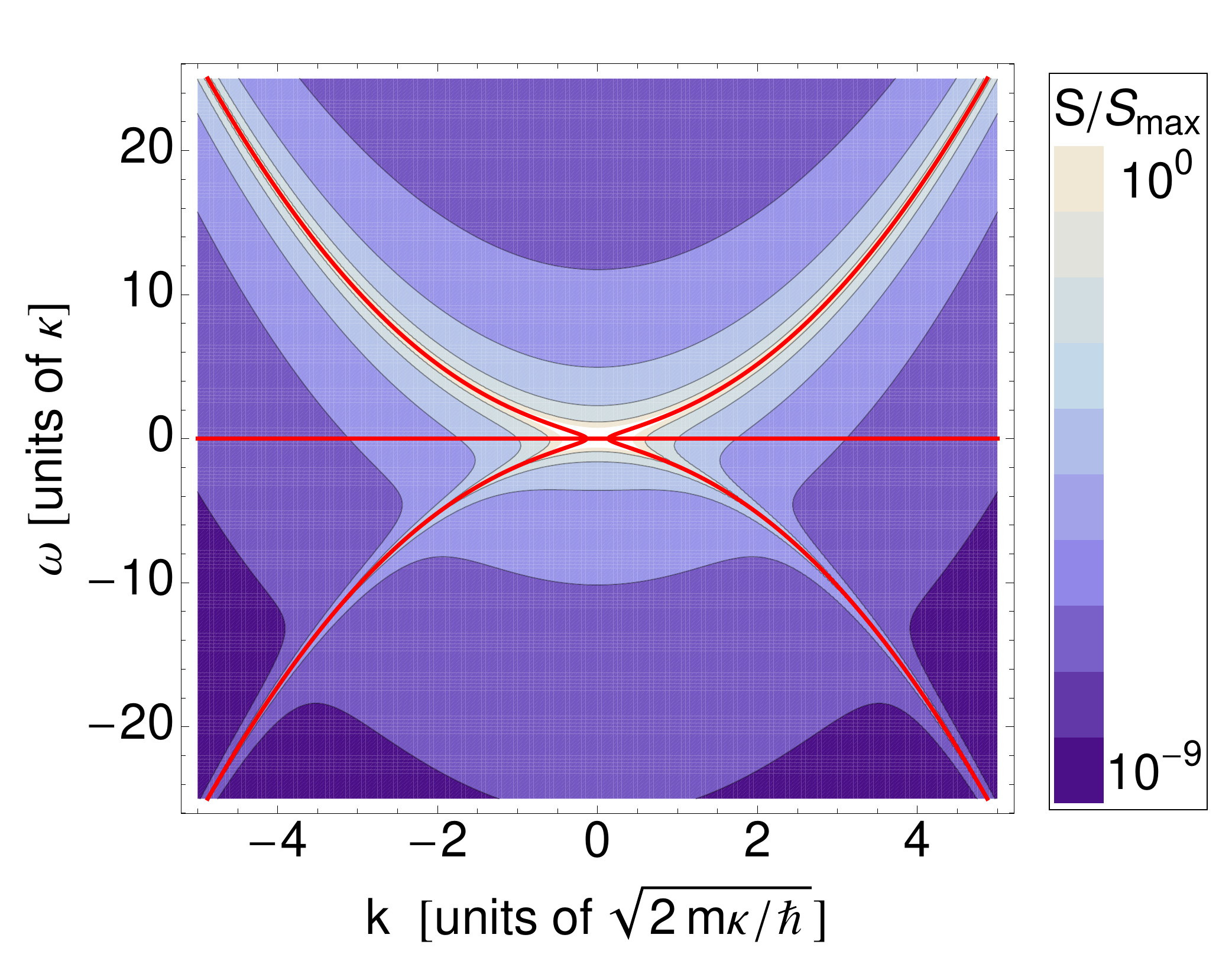}}
\caption{(Color online) Normalised momentum- and frequency-resolved spectrum of the photoluminescence from the cavity. Left panel: detuned $\nu -\omega_0 = -35 \kappa$ case with $\lambda n_A = 0$. Right panel: resonant cavity $\nu -\omega_0 = 0$ with photon-photon interactions $\lambda n_A = 0.1 \kappa$. Other system parameters: $\gamma = 10\kappa$, $g\sqrt{n_A}=7\kappa$ , $x = 7$.}
\label{fig:7}
\end{figure*}

In addition to the one-time observables discussed in the previous Section, the quantum Langevin equations also allow for a straightforward evaluation of two-time observables. In particular, we shall concentrate here in the photoluminescence spectrum,
\begin{equation}
S_\kk(\omega) =\int \frac{dt}{2\pi} \,e^{-i\omega t}\,\left\langle \bhd_\kk(t)\,\bh_\kk(0) \right\rangle
\end{equation}
which is accessible from a frequency- and angle-resolved measurement of the emission from the cavity. {A detailed study of this quantity in an equilibrium context can be found in~\cite{Marchetti2007}. A non-equilibrium calculation using linearised Keldysh techniques was reported in~\cite{SZYM2006}.}

In our quantum Langevin approach~\cite{GARD2004}, this spectrum is directly obtained as the top-left element of the matrix 
\begin{equation}
 \mathbb{S}_\kk(\omega) = \frac{1}{2\pi}(\mathbb{A}_\kk - i\omega)^{-1}\mathbb{D}(\mathbb{A}_\kk^\dagger + i\omega)^{-1}:
\end{equation}
the resonant denominators in the right-hand side of this equation shows that the photoluminescence spectrum is peaked along the real part of the Bogoliubov dispersion, while the linewidth of the peaks is set by the imaginary part.

Among the most interesting and non-trivial examples, we show in Fig.\ref{fig:7} the photoluminescence spectrum for two cases of a negative detuning $\delta<0$ (left) and of finite photon-photon interactions $\lambda>0$ (right): in both cases, photons are effectively interacting and the Bogoliubov transformation is expected to give spectral weight to the negative "ghost" branch of the Goldstone mode as well~\cite{Byrnes}. While this feature is clearly visible in the central panel, the effective interaction in the left panel is too weak to give an appreciable effect on this scale: the emitter-cavity detuning that is required for this purpose is in fact much larger than the amplification bandwidth of the emitters and therefore hardly compatible with condensation. 

At generic wavevectors and frequencies, the cavity luminescence from the dispersionless branches is typically suppressed by the detuning from the cavity mode. The only exception are the crossing points with the cavity mode, where clear peaks can be observed thanks to the resonance of the upper sideband of the Mollow triplet with the cavity mode (not shown).

\section{The Stochastic Gross-Pitaevskii equation}
\label{sec:SGPE}

\begin{figure*}[htbp]
 \subfigure[Normal ordering, $\delta =0$, $\lambda\, n_A = 0$.]{\includegraphics[width=5.5cm]{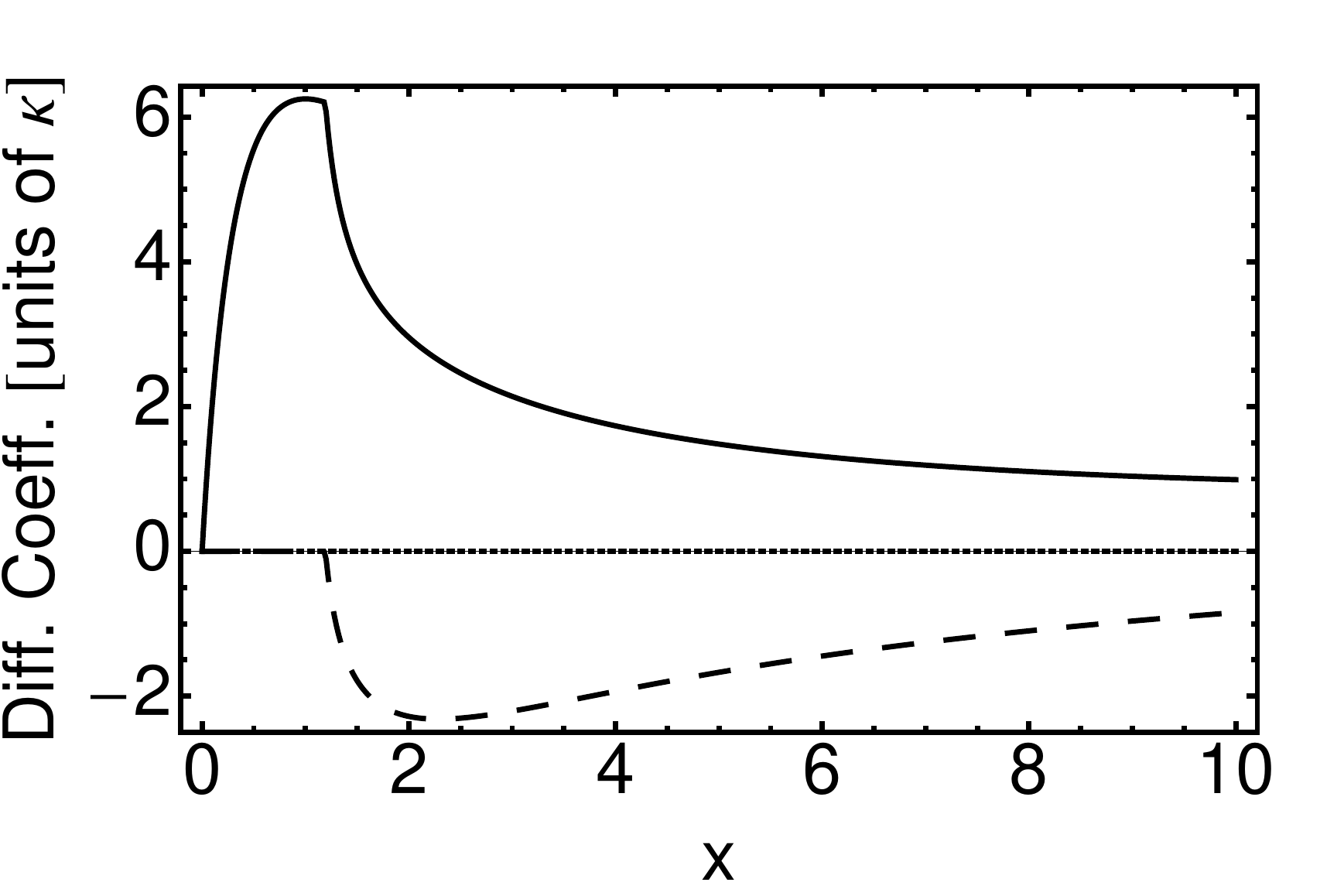}}
  \hspace{0.2mm}
  \subfigure[Normal ordering, $\delta =-150\kappa $, $\lambda\, n_A = 0$.]{\includegraphics[width=5.5cm]{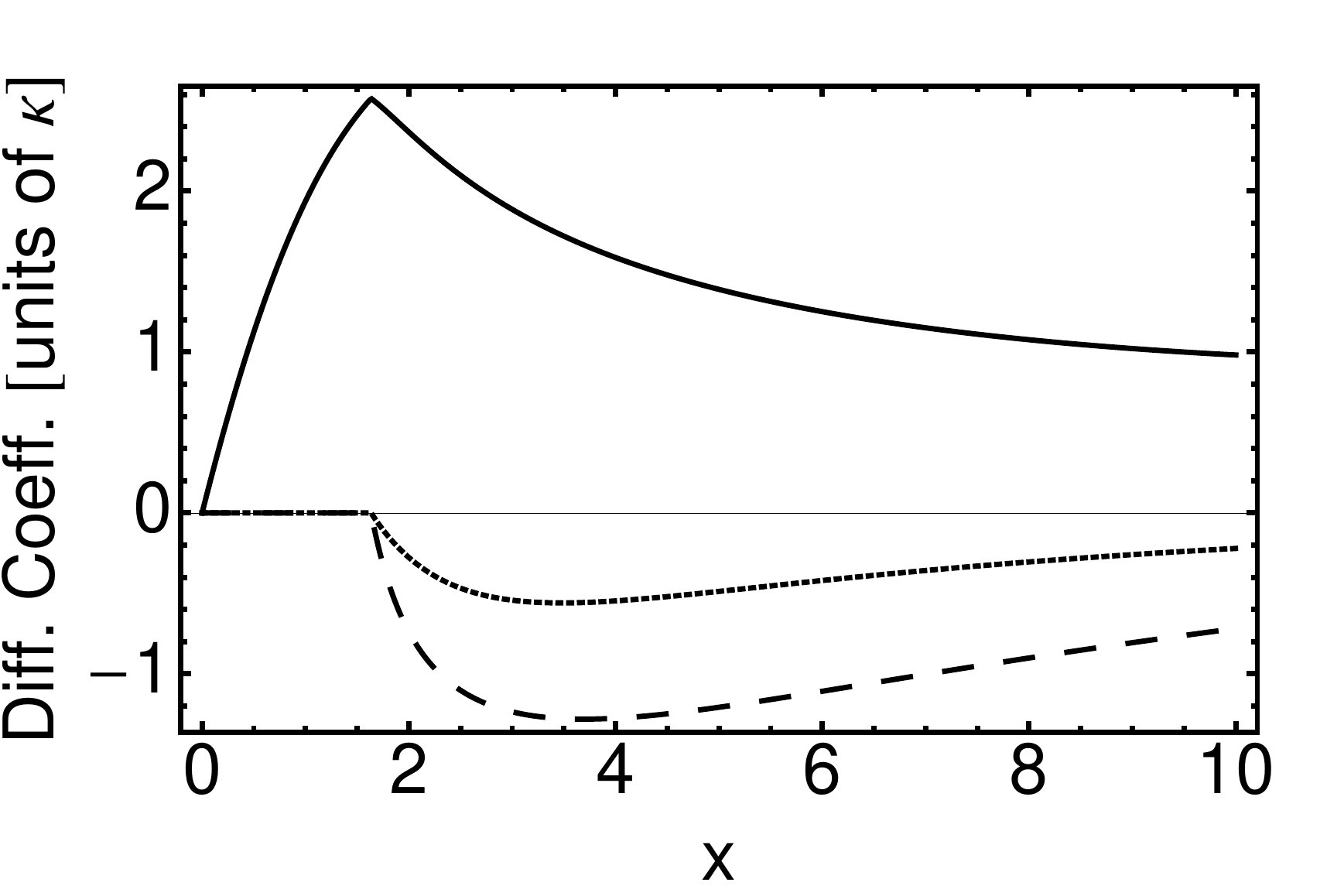}}
  \hspace{0.2mm}
\subfigure[Normal ordering, $\delta =0$, $\lambda\, n_A = 0.05\kappa$.]{ \includegraphics[width=5.5cm]{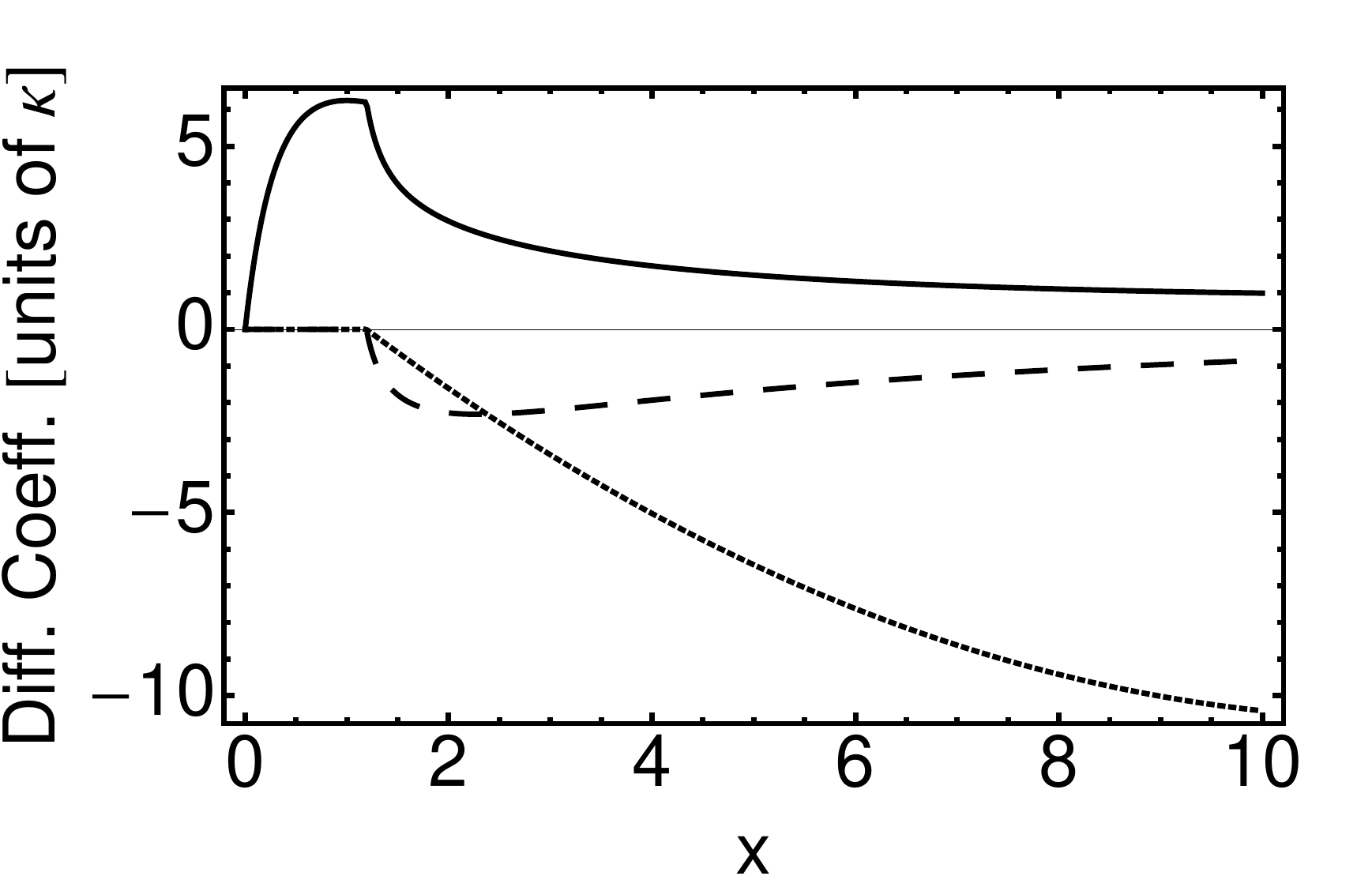}} \\
\subfigure[Wigner ordering, $\delta =0$, $\lambda\, n_A = 0$.]{\includegraphics[width=5.5cm]{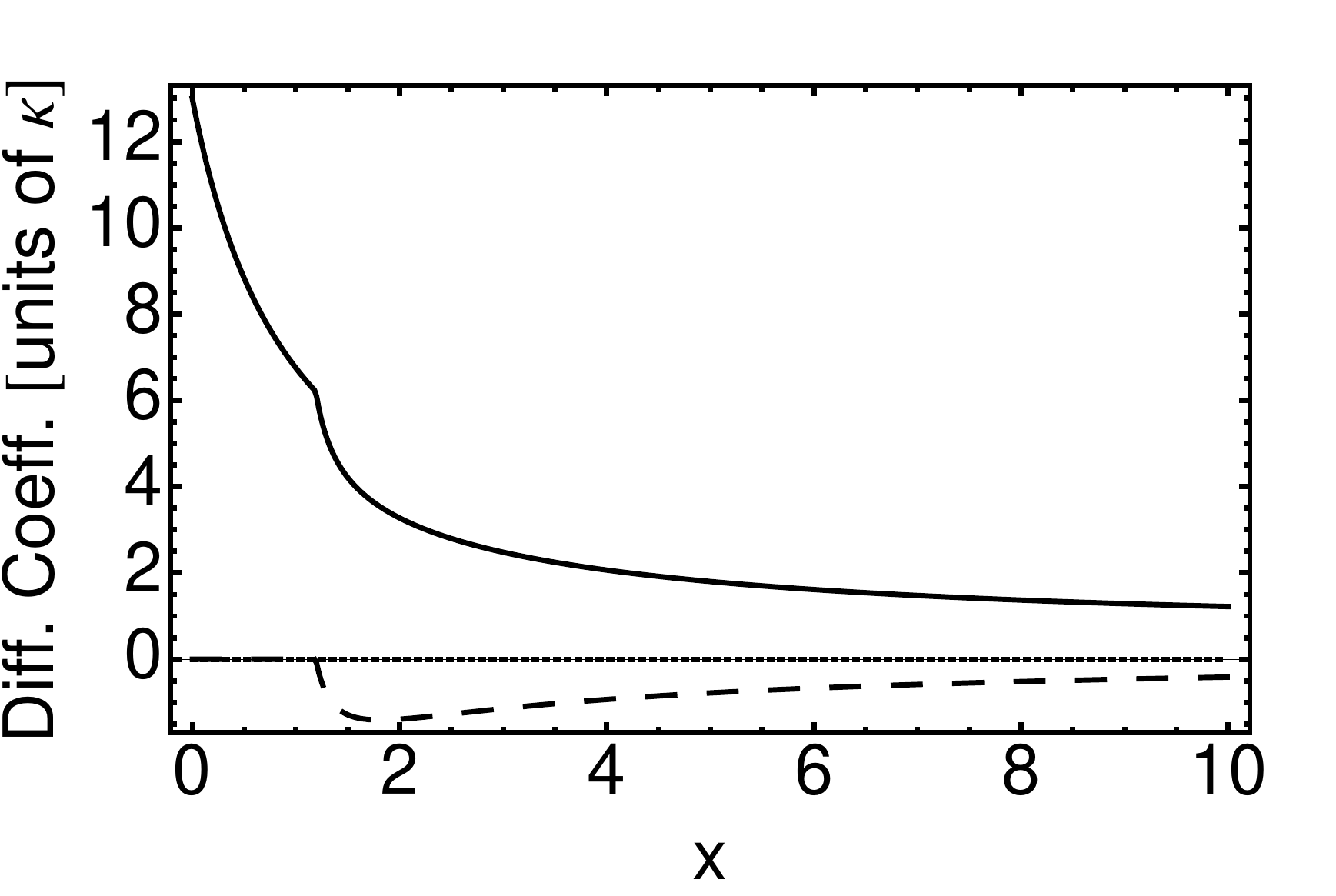}}
  \hspace{0.2mm}
  \subfigure[Wigner ordering, $\delta =-150\kappa $, $\lambda\, n_A = 0$.]{\includegraphics[width=5.5cm]{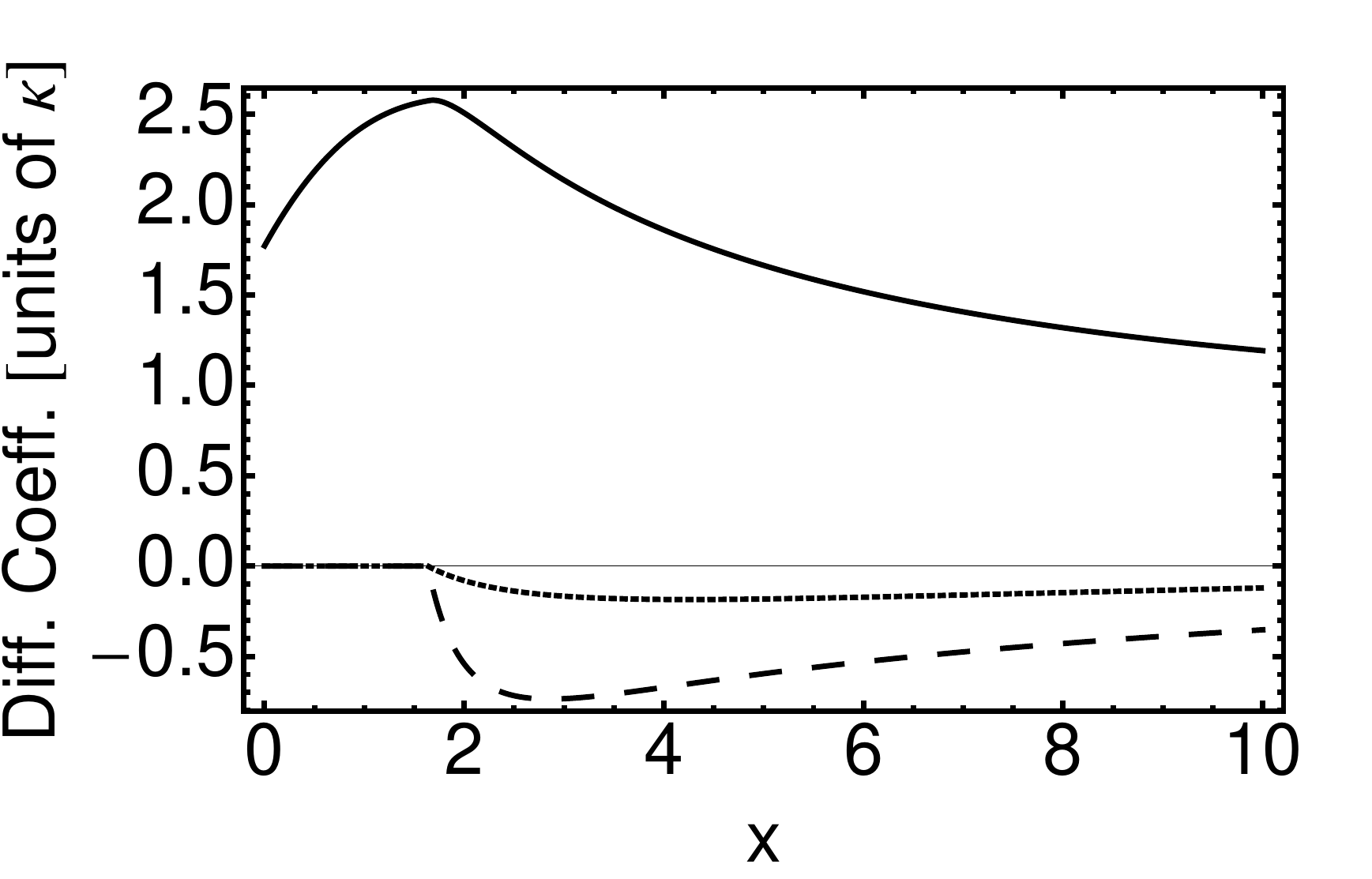}}
  \hspace{0.2mm}
\subfigure[Wigner ordering, $\delta =0$, $\lambda\, n_A = 0.05\kappa$.]{ \includegraphics[width=5.5cm]{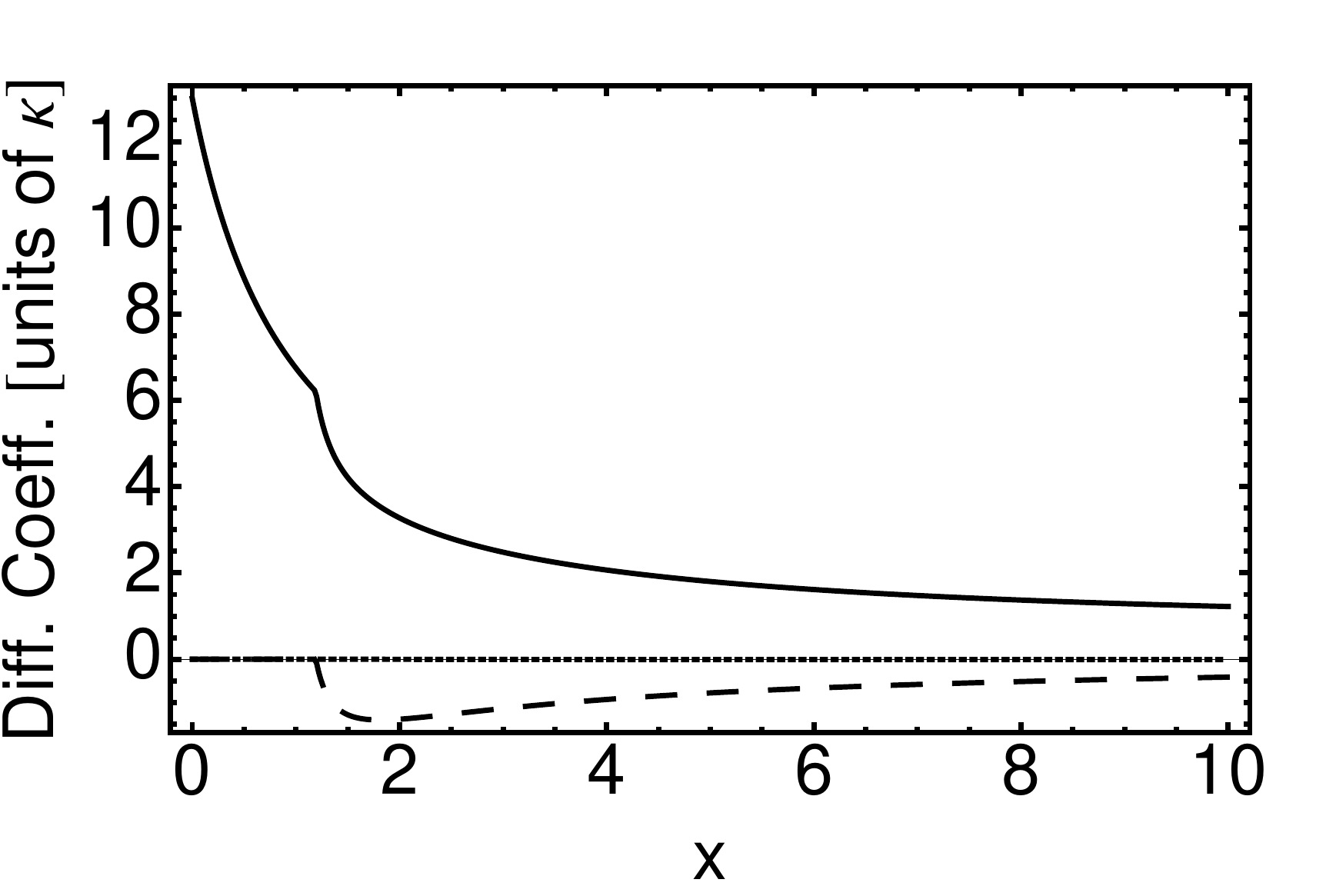}}
\caption{{Diffusion coefficients $D_{\phi\phi^*}$ (solid lines), $\text{Re}[D_{\phi\phi}]$ (dashed lines) and $\text{Im}[D_{\phi\phi}]$ (dotted lines) appearing {in the} SGPE for a field $\psi$ equal to the mean-field steady state. The quantities are plotted as a function of the pumping parameter $x = d/\gamma$ for different regimes of photon-photon interactions (left to right). The top (bottom) row refers to the SGPE in the normal (Wigner) ordering case. In all panels, we have taken $\gamma = 100 \kappa$ and $g\sqrt{n_A} = 25\kappa$. }}
\label{fig:9}
\end{figure*}
%
%

 \begin{figure*}[htbp]
 \subfigure[$\gamma = 100 \kappa$, $g\sqrt{n_A} = 25\kappa$.]{\includegraphics[width=5cm]{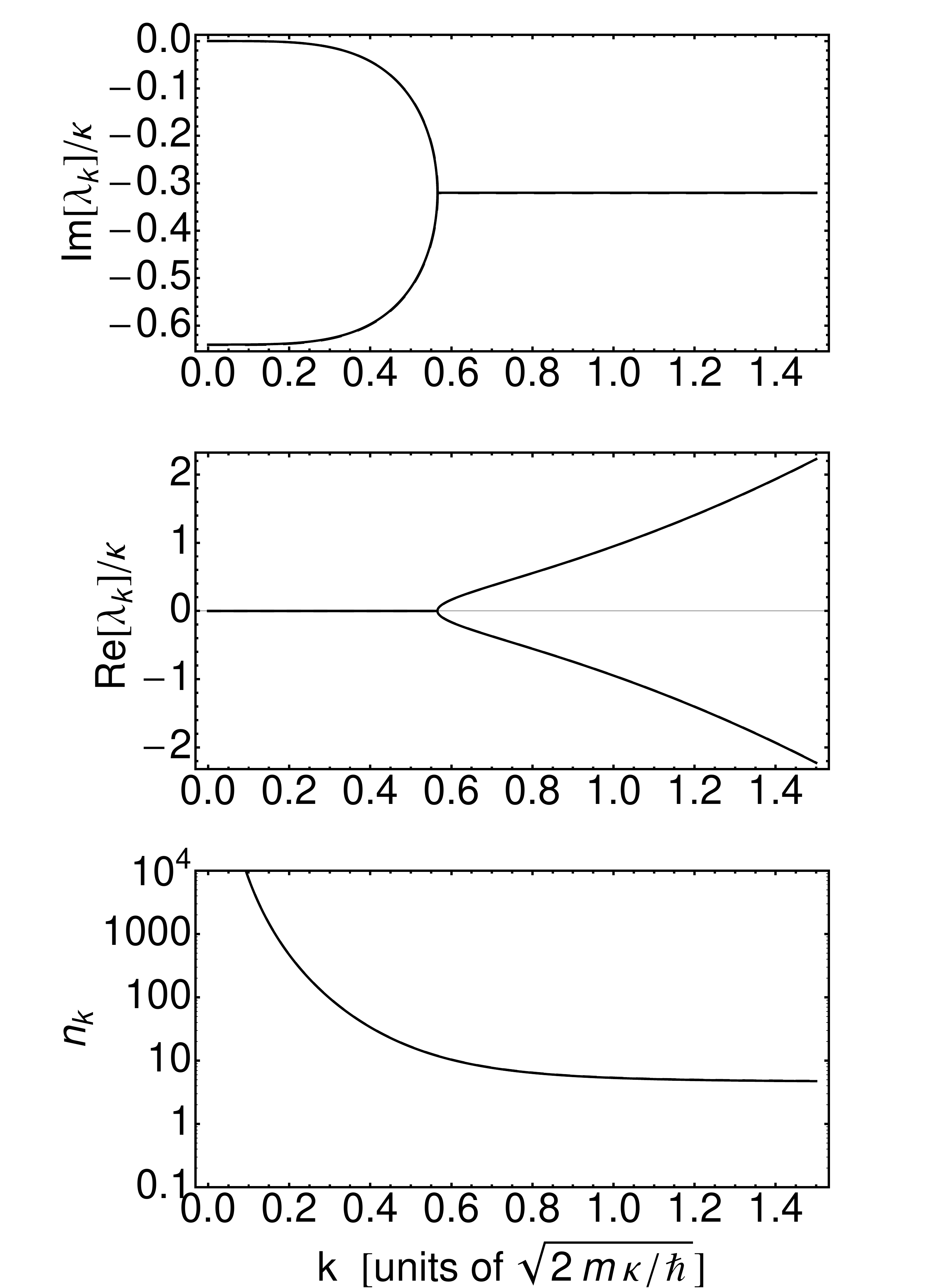}}
  \hspace{0.2mm}
  \subfigure[$\gamma = 10 \kappa$, $g\sqrt{n_A} = 7\kappa$.]{\includegraphics[width=5cm]{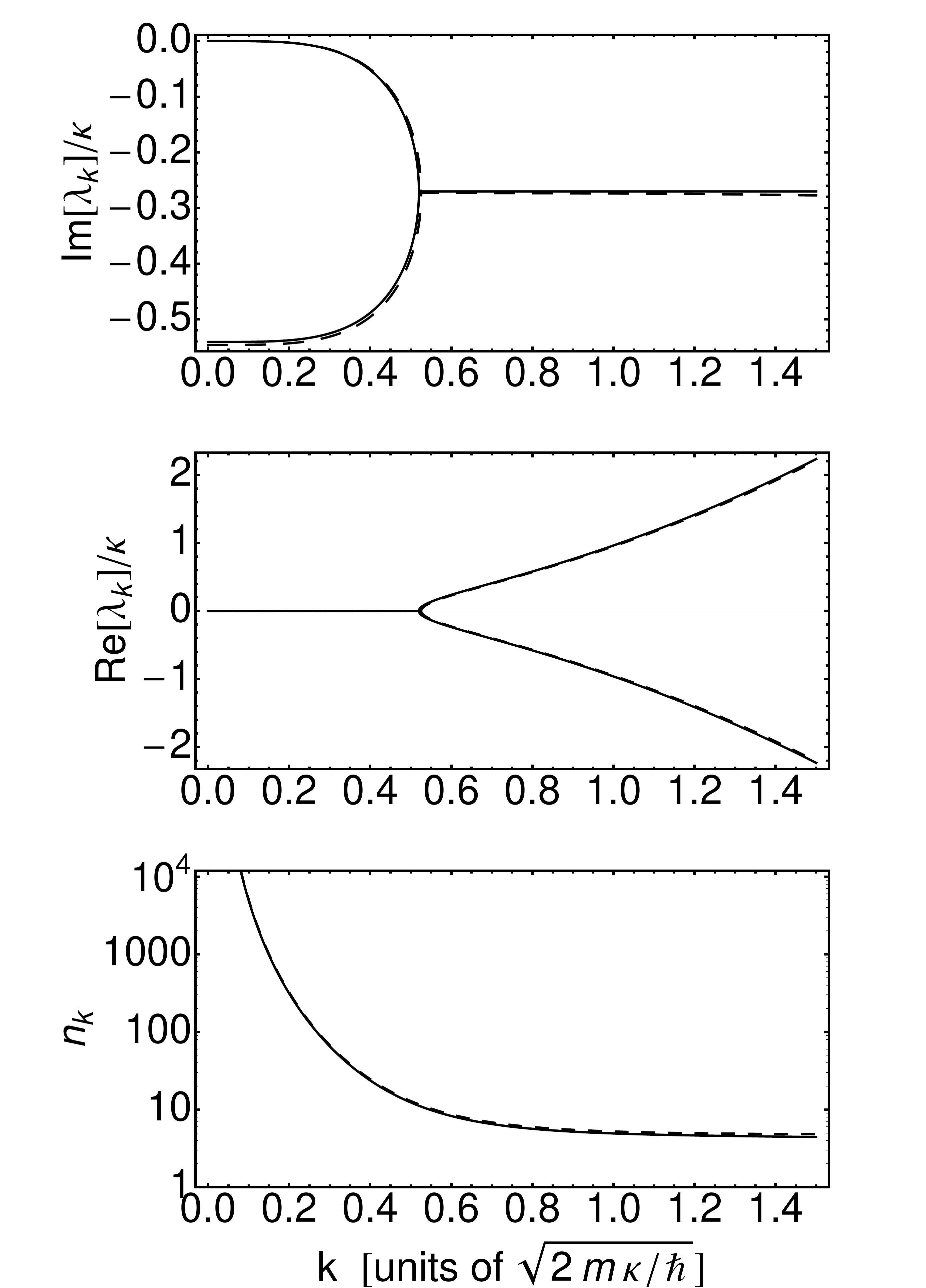}}
  \hspace{0.2mm}
\subfigure[$\gamma = 1 \kappa$, $g\sqrt{n_A} = 2.5\kappa$.]{ \includegraphics[width=5cm]{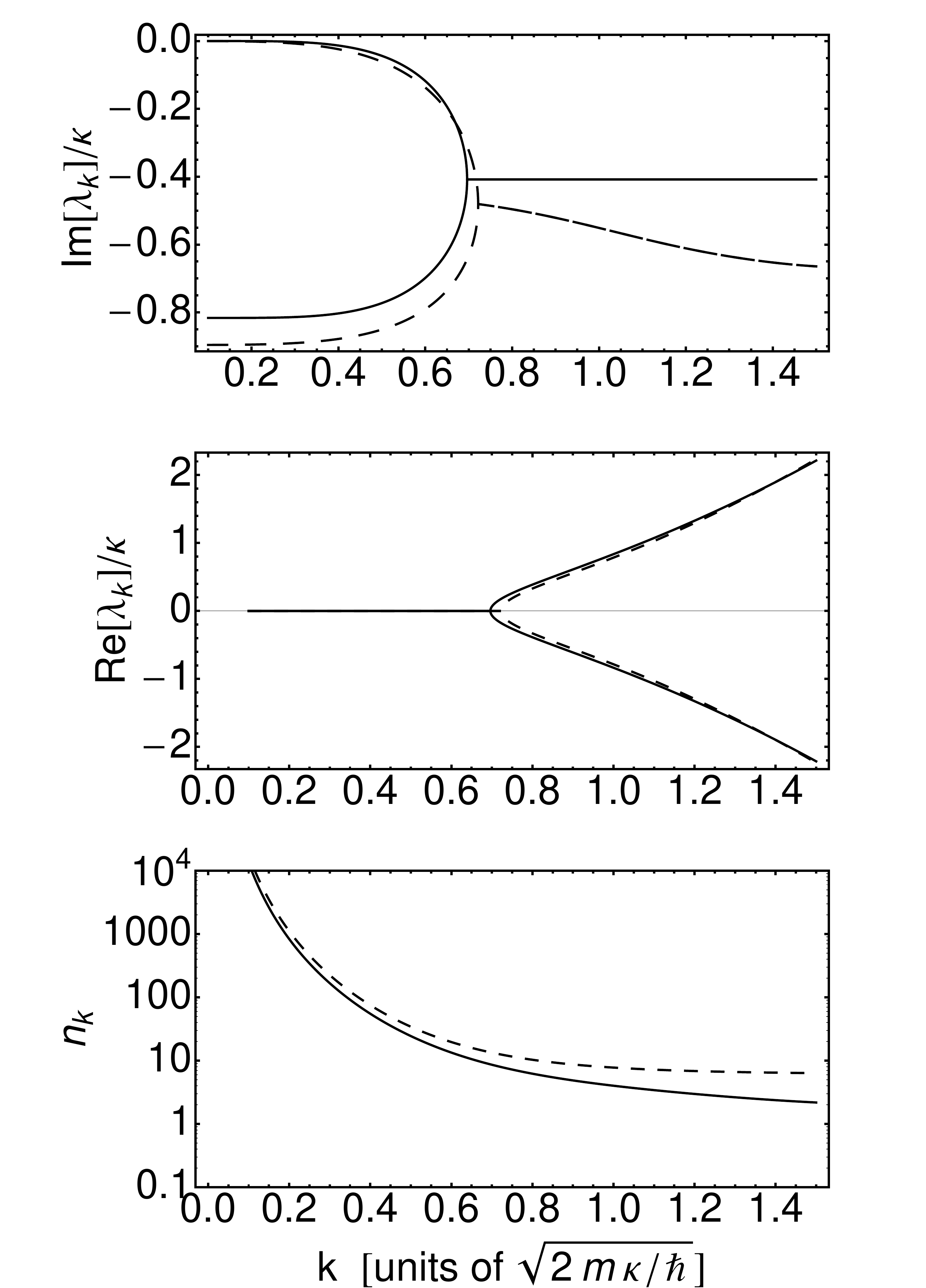}}
\caption{Comparison between SGPE the full model. First row and second row: eigenvalues of the Bogoliubov matrix in functions of the momentum; solid lines refers to SGPE quantities, dashed ones to the full model. Last row: Momentum distributions. {In all panels,} $\lambda n_A = \nu - \omega_0 =0$, $x = 2$. }
\label{fig:8}
\end{figure*}

In the previous Sections we have developed a microscopic model of condensation from which we have obtained predictions for some most interesting observable quantities. In this final Section, we are going to discuss how our model can be reduced under suitable approximations to a simpler quantum Langevin equation for the cavity field only. {In particular, we shall concentrate on the good cavity limit $\Gamma/\kappa\gg 1$, where the dynamics of the cavity field occurs on a much faster time scale as compared to the one of the emitters, which can therefore be adiabatically eliminated. Throughout this last section, we will sacrifice mathematical rigour in favor of physical intuition.} 

\subsection{{Adiabatic elimination}}

Expressing the fields in the rotating frame as:
\begin{equation}
 \phi^\dagger = \psi^\dagger\ee^{i\omega t}, \qquad S^+ = \mathcal{S}^+\ee^{i\omega t}, \qquad S^z = \mathcal{S}^z,
\end{equation}
the real-space equations of motion \eqref{eq:space1}-\eqref{eq:space3} can be rewritten as
\begin{gather}
\frac{\partial \mathcal{S}^z}{\partial t}  =  \Gamma\left(n_A\frac{\mathcal{D}}{2} - \mathcal{S}^z\right) + g\left(\mathcal{S}^+\psi + \psi\dg \mathcal{S}^-\right) + G^z ,\label{eq:adel3_1}\\
\frac{\partial \mathcal{S}^+}{\partial t}  =  -\frac{\Gamma}{2}\left(1-i\delta\right)\mathcal{S}^+ -2g\,\psi\dg \mathcal{S}^z + \widetilde{G}^+,\label{eq:adel3_2}\\
\begin{split}
\frac{\partial\psi\dg}{\partial t}  =  -\frac{\kappa}{2}\left(1+i\delta\right)\psi\dg -&  i\frac{\nabla^2}{2m}\psi\dg -g\mathcal{S}^++ \\
                                                           &+i\lambda \psi\dg\psi\dg\psi+ \widetilde{F}\dg ,\label{eq:adel3_3}
\end{split}
\end{gather}
where $\widetilde{G}^+ = \ee^{-i\omega t}G^+$ and $\widetilde{F}\dg = \ee^{-i\omega t}F\dg$. In the spirit of~\cite{LUGI1984}, the limit $\sigma\rightarrow +\infty$ can be taken provided that the quantities $ g\sqrt{n_A}/\Gamma$, $\delta$, $\langle\widetilde{G}^\alpha\widetilde{G}^{\alpha'}\rangle/n_A^2\Gamma^2$ remain finite and that the average $\lambda\langle\psi\dg\psi\rangle$ remains negligible with respect to $\Gamma$. 

{While rigourous ways to perform adiabatic elimination for ordinary differential equations exist, the situation is more complicate for our stochastic and quantum case. In what follows we shall then follow a heuristic path inspired from laser theory \cite {LAX_IX, LOUI1973} whose validity can be checked {\em a posteriori} by comparing its predictions with the full model in the linearised case; a brief discussion of a simplified but illustrative example is given in the Appendix. A rigorous derivation of the whole approach is of course needed, but goes far beyond the scope of the present work. 

As a first step, we note that time derivatives} of the spin densities can be dropped from the equations as they are negligible for large $\Gamma$. The spin operators can therefore be expressed in terms of the cavity field using the equations: 
\begin{gather}
0  =  \Gamma\left(n_A\frac{\mathcal{D}}{2} - \mathcal{S}^z\right) + g\left(\mathcal{S}^+\psi + \psi\dg \mathcal{S}^-\right) + G^z ,\label{eq:adel4_1}\\
0  =  -\frac{\Gamma}{2}\left(1-i\delta\right)\mathcal{S}^+ -2g\,\psi\dg \mathcal{S}^z + \widetilde{G}^+,\label{eq:adel4_2}\\
0  =  -\frac{\Gamma}{2}\left(1+i\delta\right)\mathcal{S}^- -2g\, \mathcal{S}^z \psi+ \widetilde{G}^-\label{eq:adel4_3}.
\end{gather}
From \eqref{eq:adel4_2} and \eqref{eq:adel4_3}, $\mathcal{S}^+$ and  $\mathcal{S}^-$ can be expressed in terms of $\mathcal{S}^z$ as
\begin{gather}
\label{eq:adel6}
S^+ = \frac{2}{\Gamma(1-i\delta)}\left( - 2 g \psi^\dagger \mathcal{S}^z + \widetilde{G}^+\right), \\
S^- = \frac{2}{\Gamma(1+i\delta)}\left( - 2 g \mathcal{S}^z \psi + \widetilde{G}^-\right), 
\end{gather}
and hence inserted in \eqref{eq:adel4_1}, which reads:
\begin{equation}
\label{eq:adel5}
\mathcal{S}^z = n_A\frac{\mathcal{D}}{2} - \frac{8g^2}{\Gamma^2(1+\delta^2)}\psi^\dagger\mathcal{S}^z\psi + \mathbb{G}^z,
\end{equation}
where 
\begin{equation}
\mathbb{G}^z = \frac{2g}{\Gamma^2(1-i\delta)}\widetilde{G}^+\psi + \frac{2g}{\Gamma^2(1+i\delta)}\psi^\dagger\widetilde{G}^- + \frac{1}{\Gamma} G^z.
\end{equation}
{While equal-time spin and cavity operators commute in the full theory, this is no longer true after the elimination, as it was noticed in \cite{LAX_IX}. An ambiguity therefore arises when writing \eqref{eq:adel4_2} and \eqref{eq:adel4_3}. In the following, inspired by~\cite{LAX1968}, we heuristically propose to choose the generalised normal ordering, $\psi^\dagger\mathcal{S}^+ \mathcal{S}^z \mathcal{S}^- \psi$. This issue is important when solving  Eq. \eqref{eq:adel5} for $\mathcal{S}^z$, which can be done by formally iterating on $\mathcal{S}^z$:}
\begin{eqnarray}
\mathcal{S}^z & = &  n_A\frac{\mathcal{D}}{2} \sum_{m=0}^{+\infty}\frac{(-1)^m}{n_s^m}\left(\psi^\dagger\right)^m\psi^m +\sum_{m=0}^{+\infty}\frac{(-1)^m}{n_s^m}\left(\psi^\dagger\right)^m\mathbb{G}^z\psi^m \nonumber \\
& = &  n_A\frac{\mathcal{D}}{2} :\frac{1}{1+\frac{\psi^\dagger\psi}{n_s}}: + :\frac{1}{1+\frac{\psi^\dagger\psi}{n_s}}\mathbb{G}^z:,
\end{eqnarray}
where columns denote normal ordering and the saturation density is defined as
\begin{equation}
n_s=\frac{\Gamma^2}{8g^2}(1+\delta^2).
\end{equation}
The explicit expression for $\mathcal{S}^z$ can be inserted back in \eqref{eq:adel6} to obtain the expression for $\mathcal{S}^+$ and  $\mathcal{S}^-$, which can be finally substituted in \eqref{eq:adel3_3} to give a quantum stochastic Gross-Pitaevskii equation
\begin{equation}
\label{SGPEop}
\begin{split}
 \frac{\partial \psi\dg}{\partial t} = -\frac{\kappa}{2}(1+i\delta)\psi\dg & - i\frac{\nabla^2}{2m}\psi\dg +\\
  &+\psi\dg:\frac{P_0(1+i\delta)}{1+\frac{\psi\dg\psi}{n_s}}: + i\lambda \psi\dg\psi\dg\psi +\mathbb{F}\dg,
  \end{split}
\end{equation}
where the pumping coefficient has the form
\begin{equation}
P_0 = \frac{2g^2n_A\mathcal{D}}{\Gamma(1 + \delta^2)}, 
\end{equation}
and $\mathbb{F}\dg$ is a new effective noise operator given by
\begin{equation}
\mathbb{F}\dg = \widetilde{F}\dg -\frac{2g}{\Gamma(1-i\delta)}\widetilde{G}^+ + \frac{4g^2}{\Gamma(1-i\delta)}:\psi^\dagger\frac{1}{1+\frac{\psi^\dagger\psi}{n_s}}\mathbb{G}^z:.
\end{equation}
The diffusion matrix of the noise $\mathbb{F}\dg$ depends on the field state $\psi$ and $\psi^\dagger$ and can be written in the form
\begin{equation}
\left(
\begin{array}{cc}
\langle\mathbb{F}\dg(\xx,t)\mathbb{F}(\xx',t')\rangle & \langle\mathbb{F}\dg(\xx,t)\mathbb{F}\dg(\xx',t')\rangle \\ 
\langle\mathbb{F}(\xx,t)\mathbb{F}(\xx',t')\rangle & \langle\mathbb{F}(\xx,t)\mathbb{F}\dg(\xx',t')\rangle 
\end{array}
\right)
=
\left(
\begin{array}{cc}
A & C^* \\
C & B
\end{array}
\right),
\end{equation}
where $A$, $B$, $C$ are functions of $\psi$ and $\psi^\dagger$. Note in particular the non-zero $C$ term in the non-diagonal positions, which originates from the contribution of the  emitter noise operators $G^\alpha$ ($\alpha = +,-,z$) to the resulting noise $\mathbb{F}$. 

\subsection{{Normally-ordered c-number representation}}
{A useful technique to obtain physical predictions from the operator-valued stochastic Gross-Pitaevskii \eqref{SGPEop}, is to represent it in terms of} an equivalent c-number equation. In doing this, we follow the procedure explained in~\cite{Benkert}. {As one typically does for phase-space representations~\cite{GARD2000}, the first step is to choose an ordering prescription for the operator products according to which all quantities of the theory have to be consistently expressed.}

A first choice is to assume normal ordering. In this case, the operator-valued SGPE \eqref{SGPEop} gets projected onto the c-number Ito SGPE
\begin{equation}
\label{SGPE}
\begin{split}
 i\mathrm{d}\psi = &\left[\omega_0 -\frac{\nabla^2}{2m} +  \right. \lambda|\psi|^2 +\frac{P_0\delta}{1+\frac{|\psi|^2}{n_s}}+ \\
 &+\left.i\left(\frac{P_0}{1+\frac{|\psi|^2}{n_s}}-\frac{\kappa}{2}\right)  \right]\psi \,\mathrm{d}t + \mathrm{d}W.
 \end{split}
\end{equation}
A similar equation was derived in the early theory of laser \cite{HAKE1975}. {The second order momenta of the noise} have local spatial and temporal correlations
\begin{eqnarray}
 \left \langle \dd W(\mathbf{x},t)\dd W^*(\mathbf{x'},t)   \right \rangle & = & 2D_{\psi\psi^*}(\xx)\,\delta^{(d)}(\mathbf{x} - \mathbf{x'})\dd t, \\
 \left \langle \dd W(\mathbf{x},t)\dd W(\mathbf{x'},t)     \right \rangle & = & 2D_{\psi\psi}(\xx)\,\delta^{(d)}(\mathbf{x} - \mathbf{x'})\dd t
\end{eqnarray}
{and} their variances $D_{\psi\psi^*}(\xx)$ and $D_{\psi\psi}(\xx)$ depend locally on the field $\psi(\xx)$. Their value can be determined by imposing that the motion equation for the second moments of the field determined by the c-number equation \eqref{SGPE} must be equal to the ones obtained from the operatorial equation \eqref{SGPEop} in the normal ordered form. Using this prescription, we obtain:
 \begin{equation}
 \begin{cases}
 2D_{\psi\psi^*} = A,\\
 2D_{\psi\psi} = C - \frac{P_0(1-i\delta)}{\left(1+\frac{|\psi|^2}{n_s}\right)^2} \frac{\psi^2}{n_s} -i\lambda\psi^2. \label{psipsi}
 \end{cases}
 \end{equation} 
As expected from the U(1) symmetry of the original problem,  both $C$ and the normal ordering terms in \eqref{psipsi} are all proportional to $\psi^2$. The dependence of the diffusion coefficients on the pumping parameter $x=d/\gamma$ are plotted in Fig. \ref{fig:9} for the mean-field steady state. Remarkably, while $D_{\psi^*\psi}$ and $\text{Re}[D_{\psi\psi}]$ depend very slowly on $x$ and are not much affected by the presence of detuning or self-interaction, the imaginary part $\text{Im}[D_{\psi\psi}]$ crucially depends on these parameters. Note that the possibility of a non-vanishing $D_{\psi\psi}$ variance was overlooked in the phenomenological discussion that we published in~\cite{ACIC2013} {and has not been taken into account in~\cite{Sieberer2013,Altman,Wouters2013}.}

{Due to the saturable pumping term in the SGPE, higher-order momenta of the noise are present beyond the usual Gaussian noise. Their correlation can be extracted by considering the equation of motion for higher-order operator products. Inspired by the so-called truncated Wigner scheme~\cite{Ciuti2005,RMP2013}, one can expect that their contribution is actually negligible in the mean-field limit discussed in Sec.\ref{sec:Fluctuations}.}

\subsection{{Comparison with full calculation}}
As a check of the validity of this reformulation, in Fig.\ref{fig:8} we compare the predictions of the SGPE for the dispersion of the collective Bogoliubov modes (upper and central row) and for the momentum distribution (lower row) with the predictions of the full model as derived in Sec.\ref{sec:Fluctuations}. 

The Bogoliubov dispersion is obtained by linearising the deterministic part of the SGPE equation \eqref{SGPE} around the steady-state. a straightforward calculation gives a dispersion analogous to the one originally obtained in~\cite{WOUT2007},
\begin{equation}
\omega^\pm_\kk = -\Gamma_p \pm \sqrt{\Gamma_p^2 - E_\kk^2}
\end{equation} 
with the damping parameter $\Gamma_p= \kappa(2P_0 -\kappa)/4P_0$ and the equilibrium Bogoliubov dispersion $E_\kk= \sqrt{\epsilon_\kk(\epsilon_\kk + 2\lambda_{\rm eff}|\beta_0|^2)}$. In this latter, note that the effective nonlinear term
\begin{equation}
\lambda_{\rm eff} = \lambda - \frac{\kappa}{2}\frac{\delta}{n_s + |\beta_0|^2},
\end{equation} 
contains two contribution: the former results from the direct photon-photon interaction $\lambda$, the latter describes the effective Kerr optical nonlinearity due to saturation of the emitters~\cite{COHE2004}.

The momentum distribution shown in the bottom row is instead obtained by reintroducing the noise terms in the linearised equation and then making a small noise expansion: average of fluctuation operators like $n^s_\kk$ are written as a linear function of the noise variances $D_{\psi^*\psi}$ and $D_{\psi\psi}$. 
 
In the three columns of Fig.\ref{fig:8}, we show the result of the comparison for different system parameters: as one moves deeper in the good cavity limit (left panels), the agreement becomes very good, while significant discrepancies are expected outside this limit (right panels). As expected, the adiabatic elimination procedure for the momentum distribution is only valid at sufficiently low $k$ when the cavity field detuning is small as compared to the atomic linewidth: breakdown of this condition is indeed visible in the bottom-right panel, where a clear qualitative deviation appears at large $k$. In particular, the adiabatic elimination of the emitters in the SGPE loses track of frequency dependence amplification and therefore is not able to recover the large $k$ behaviour of the momentum distribution. Note also that the quantitative agreement visible in the figure crucially relies on the correct inclusion of the $D_{\psi\psi}$ variance. 

In spite of its accurate predictions illustrated in Fig.\ref{fig:8}, the stochastic equation \eqref{SGPE} is only meaningful at a linearised level. A closer look at {the top row of} Fig.\ref{fig:9} shows in fact that $|D_{\psi\psi}|$ is not always lower or equal to $D_{\psi\psi^*}$, as it is expected from the Cauchy-Schwartz inequality for a generic Ito stochastic equation~\cite{GARD2004}. While at the linearised level one can forget this fact and formally solve the linear stochastic equation irrespectively on the positivity of the noise variance, this is no longer possible when one wishes to describe the nonlinear dynamics stemming from large fluctuations, e.g. in the vicinity of the critical point for condensation. This feature, often neglected in laser theory \cite{LOUI1973}, is particularly visible in the interacting case for $\lambda\neq 0$ or $ \delta \neq 0$. Techniques for numerically solving (generalised) stochastic differential equations with non-positive-definite noise were proposed, {the best known example being the so-called Positive-P representation which however keeps suffering from other difficulties~\cite{GARD2000}.}

\subsection{{Symmetrically-ordered c-number representation}}

Another possible way-out is to make a different choice for the ordering of operators when performing the projection of the operator-valued SGPE \eqref{SGPEop} onto the c-number SGPE{, e.g. the symmetric ordering of Wigner representation where c-number averages correspond to symmetrically ordered quantities. In this case, the variance matrix of the noise is indeed positive definite {(see bottom row of Fig.\ref{fig:9})}, but several other difficulties appear~\cite{GARD2000,RMP2013}. Firstly, the normal ordered saturation term in Eq. \eqref{SGPEop} cannot be easily symmetrised, which complicates writing of the deterministic part of the stochastic equation.} Secondly, the symmetrisation of any non-linear term in \eqref{SGPEop} {produces} terms proportional to the commutator $[\psi(\xx),\psi^\dagger(\xx)]$, which is a {UV} divergent quantity. Finally, any non linear term in \eqref{SGPEop} will generate a noise {with non-vanishing third order momenta, e.g. $\langle \dd W^2 \dd W^*\rangle \propto \dd t$.}

The first two problems can be overcome: the saturation term can be approximated truncating the power expansion to some order, so that symmetrisation becomes viable. {A finite expression for the field commutator is available if one discretises the field on a lattice, which corresponds to broadening the delta-function according to the smallest accessible length-scale of the system.} The third problem poses a more challenging task, as noise with such features is extremely difficult to treat. {Solutions have been proposed~\cite{PLIMAK,Polkovnikov} but never implemented into the simulation of large systems. Note that this is a well-known issue in the theory of phase-space representation of quantum fields, where interaction terms generate third-order derivatives in the equation for the Wigner function, spoiling its interpretation as a Fokker-Planck equation~\cite{GARD2000,VogelRisken}. As already mentioned, truncated-Wigner simulations where these terms are neglected are expected to be correct in the mean-field limit and have been used in simulations of polariton condensation in~\cite{Ciuti2005}.}
\section{Conclusions}
\label{sec:Conclusions}

In this Article, we have built on top of laser theory to develop a quantum {field} model of {non-equilibrium} Bose-Einstein condensation of photons/polaritons in {planar} microcavity devices. The system under examination consists of a spatially extended cavity mode coupled to a continuous distribution of externally pumped two-level emitters and is described in terms of quantum Langevin equations. {In our view, this is a minimal model that is able to describe non-equilibrium condensation simultaneously including at a quantum level the spatial dynamics of the cavity field, a saturation mechanism, and some frequency-dependence of the gain. We expect that such a model may become an essential tool in view of full numerical simulations of the non-equilibrium phase transition.}

As a first example of application of our theory, we have worked out the main characteristics of quantum fluctuations around the condensate state. Our calculations confirm the non-equilibrium features that were anticipated by previous theories and/or observed in the experiments: in particular, the collective Bogoliubov modes include a Goldstone branch with diffusive properties, photoluminescence is visible on both upper and lower branches of the Bogoliubov spectrum, and the momentum distribution shows a large-$k$ decrease even in the absence of any collisional thermalisation mechanism. 
{This result {provides a theoretical explanation to} the experimental {observation~\cite{BAJO2007}} that {a condensate} can exhibit thermal-like features in the momentum distribution even in the absence of thermalising collisions. Given the qualitatively different {shape of the collective excitation dispersion}, we expect that a decisive insight {in the equilibrium vs. non-equilibrium} nature of {a condensation process can} be obtained by measuring dispersions from the luminescence spectra or via pump-and-probe spectroscopy \cite{Marchetti2007, Byrnes, WOUT2010, Nyman}.}

In the good-cavity limit, {we propose a reformulation of our theory} in terms of a stochastic Gross-Pitaevskii equation. In addition to {contributing to the justification of} a widely used model of non-equilibrium statistical mechanics, this connection allows to {relate the phenomenological parameters of the SGPE to a more fundamental theory}. In particular, it turns out that the noise term originates from a complex interplay between pumping and interactions and, in some cases, can even {exhibit a multiplicative} dependence on the field. This unexpected fact may turn out to have important consequences {on the critical properties. To reliably simulate this physics in large systems, further work is needed to overcome subtle issues related to the peculiar statistics of the noise terms.}

\begin{acknowledgments}
We are grateful to Andrea Gambassi and Michiel Wouters for continuous stimulating discussions. IC acknowledges partial financial support from ERC through the QGBE grant and from Provincia Autonoma di Trento.

\end{acknowledgments}    

\appendix*
\section{Adiabatic elimination}
\label{appendix}

In this Appendix we will work out a simple example to give a more solid ground to the adiabatic elimination of Sec. \ref{sec:SGPE}.
Let us consider the following simple Ito equations
\begin{equation}
\label{eq:toymodel}
\begin{cases}
\dd x = (-\gamma\, x - g\, y )\dd t + \dd W_x, \\
\dd y = (-\Gamma\, y - g \,x )\dd t + \dd W_y.
\end{cases}
\end{equation}
Assuming to be interested in the slow function $x(t)$ in the limit of $\Gamma \gg \gamma, g $, one can formally explicit

\begin{equation}
y(t) = -g \int_{-\infty}^t \dd t' \, \ee^{-\Gamma(t-t')} x(t')  + \int_{-\infty}^t  \, \ee^{-\Gamma(t-t')} \dd W_y(t') 
\end{equation} 
and substitute its expression in the equation for $x$, to obtain
\begin{equation}
\label{eq:xexact}
\dd x = \left [-\gamma x  + g^2 \int_{-\infty}^t \dd t' \, \ee^{-\Gamma(t-t')} x(t') \right]\dd t + \dd \widetilde{W}_x,
\end{equation}
where we considered the initial time $t_0=-\infty$ and 
\begin{equation}
\label{eq:dWexact}
\dd\widetilde{W}_x = \dd W_x -g\int_{-\infty}^t  \ee^{-\Gamma(t-t')} \dd W_y(t') . 
\end{equation}
Eq. \eqref{eq:xexact} is exact and notice that $\dd\widetilde{W}_x$ now has a frequency-dependent spectrum. If $\gamma \ll \Gamma$, the kernel $\exp[-\Gamma |t|]$ has a support which is much smaller than the time-scale on which $x(t)$ varies appreciably . Therefore one can approximate it as a delta-function
\begin{equation}
\frac{\Gamma}{2}\ee^{-\Gamma|t|} \simeq \delta(t),
\end{equation}
and \eqref{eq:xexact}, \eqref{eq:dWexact} become
\begin{equation}
\begin{cases}
\dd x = - \left (\gamma  - \frac{g^2}{\Gamma} \right)x\, \dd t + \dd \widetilde{W}_x, \\
\dd\widetilde{W}_x = \dd W_x -\frac{g}{\Gamma}\dd W_y.
\end{cases}
\end{equation}
This equations are the same we would have obtained simply dropping the temporal derivative $d y/ d t$ in \eqref{eq:toymodel}, as we did in Sec. \ref{sec:SGPE}.

\bibliography{biblio}

\begin{thebibliography}{66}%
\makeatletter
\providecommand \@ifxundefined [1]{%
 \@ifx{#1\undefined}
}%
\providecommand \@ifnum [1]{%
 \ifnum #1\expandafter \@firstoftwo
 \else \expandafter \@secondoftwo
 \fi
}%
\providecommand \@ifx [1]{%
 \ifx #1\expandafter \@firstoftwo
 \else \expandafter \@secondoftwo
 \fi
}%
\providecommand \natexlab [1]{#1}%
\providecommand \enquote  [1]{``#1''}%
\providecommand \bibnamefont  [1]{#1}%
\providecommand \bibfnamefont [1]{#1}%
\providecommand \citenamefont [1]{#1}%
\providecommand \href@noop [0]{\@secondoftwo}%
\providecommand \href [0]{\begingroup \@sanitize@url \@href}%
\providecommand \@href[1]{\@@startlink{#1}\@@href}%
\providecommand \@@href[1]{\endgroup#1\@@endlink}%
\providecommand \@sanitize@url [0]{\catcode `\\12\catcode `\$12\catcode
  `\&12\catcode `\#12\catcode `\^12\catcode `\_12\catcode `\%12\relax}%
\providecommand \@@startlink[1]{}%
\providecommand \@@endlink[0]{}%
\providecommand \url  [0]{\begingroup\@sanitize@url \@url }%
\providecommand \@url [1]{\endgroup\@href {#1}{\urlprefix }}%
\providecommand \urlprefix  [0]{URL }%
\providecommand \Eprint [0]{\href }%
\providecommand \doibase [0]{http://dx.doi.org/}%
\providecommand \selectlanguage [0]{\@gobble}%
\providecommand \bibinfo  [0]{\@secondoftwo}%
\providecommand \bibfield  [0]{\@secondoftwo}%
\providecommand \translation [1]{[#1]}%
\providecommand \BibitemOpen [0]{}%
\providecommand \bibitemStop [0]{}%
\providecommand \bibitemNoStop [0]{.\EOS\space}%
\providecommand \EOS [0]{\spacefactor3000\relax}%
\providecommand \BibitemShut  [1]{\csname bibitem#1\endcsname}%
\let\auto@bib@innerbib\@empty
\bibitem [{\citenamefont {Stevenson}\ \emph {et~al.}(2000)\citenamefont
  {Stevenson}, \citenamefont {Astratov}, \citenamefont {Skolnick},
  \citenamefont {Whittaker}, \citenamefont {Emam-Ismail}, \citenamefont
  {Tartakovskii}, \citenamefont {Savvidis}, \citenamefont {Baumberg},\ and\
  \citenamefont {Roberts}}]{STEVENSON2000}%
  \BibitemOpen
  \bibfield  {author} {\bibinfo {author} {\bibfnamefont {R.~M.}\ \bibnamefont
  {Stevenson}}, \bibinfo {author} {\bibfnamefont {V.~N.}\ \bibnamefont
  {Astratov}}, \bibinfo {author} {\bibfnamefont {M.~S.}\ \bibnamefont
  {Skolnick}}, \bibinfo {author} {\bibfnamefont {D.~M.}\ \bibnamefont
  {Whittaker}}, \bibinfo {author} {\bibfnamefont {M.}~\bibnamefont
  {Emam-Ismail}}, \bibinfo {author} {\bibfnamefont {A.~I.}\ \bibnamefont
  {Tartakovskii}}, \bibinfo {author} {\bibfnamefont {P.~G.}\ \bibnamefont
  {Savvidis}}, \bibinfo {author} {\bibfnamefont {J.~J.}\ \bibnamefont
  {Baumberg}}, \ and\ \bibinfo {author} {\bibfnamefont {J.~S.}\ \bibnamefont
  {Roberts}},\ }\href {\doibase 10.1103/PhysRevLett.85.3680} {\bibfield
  {journal} {\bibinfo  {journal} {Phys. Rev. Lett.}\ }\textbf {\bibinfo
  {volume} {85}},\ \bibinfo {pages} {3680} (\bibinfo {year}
  {2000})}\BibitemShut {NoStop}%
\bibitem [{\citenamefont {Baumberg}\ \emph {et~al.}(2000)\citenamefont
  {Baumberg}, \citenamefont {Savvidis}, \citenamefont {Stevenson},
  \citenamefont {Tartakovskii}, \citenamefont {Skolnick}, \citenamefont
  {Whittaker},\ and\ \citenamefont {Roberts}}]{BAUMBERG2000}%
  \BibitemOpen
  \bibfield  {author} {\bibinfo {author} {\bibfnamefont {J.~J.}\ \bibnamefont
  {Baumberg}}, \bibinfo {author} {\bibfnamefont {P.~G.}\ \bibnamefont
  {Savvidis}}, \bibinfo {author} {\bibfnamefont {R.~M.}\ \bibnamefont
  {Stevenson}}, \bibinfo {author} {\bibfnamefont {A.~I.}\ \bibnamefont
  {Tartakovskii}}, \bibinfo {author} {\bibfnamefont {M.~S.}\ \bibnamefont
  {Skolnick}}, \bibinfo {author} {\bibfnamefont {D.~M.}\ \bibnamefont
  {Whittaker}}, \ and\ \bibinfo {author} {\bibfnamefont {J.~S.}\ \bibnamefont
  {Roberts}},\ }\href {\doibase 10.1103/PhysRevB.62.R16247} {\bibfield
  {journal} {\bibinfo  {journal} {Phys. Rev. B}\ }\textbf {\bibinfo {volume}
  {62}},\ \bibinfo {pages} {R16247} (\bibinfo {year} {2000})}\BibitemShut
  {NoStop}%
\bibitem [{\citenamefont {Baas}\ \emph {et~al.}(2006)\citenamefont {Baas},
  \citenamefont {Karr}, \citenamefont {Romanelli}, \citenamefont {Bramati},\
  and\ \citenamefont {Giacobino}}]{BAAS2006}%
  \BibitemOpen
  \bibfield  {author} {\bibinfo {author} {\bibfnamefont {A.}~\bibnamefont
  {Baas}}, \bibinfo {author} {\bibfnamefont {J.-P.}\ \bibnamefont {Karr}},
  \bibinfo {author} {\bibfnamefont {M.}~\bibnamefont {Romanelli}}, \bibinfo
  {author} {\bibfnamefont {A.}~\bibnamefont {Bramati}}, \ and\ \bibinfo
  {author} {\bibfnamefont {E.}~\bibnamefont {Giacobino}},\ }\href {\doibase
  10.1103/PhysRevLett.96.176401} {\bibfield  {journal} {\bibinfo  {journal}
  {Phys. Rev. Lett.}\ }\textbf {\bibinfo {volume} {96}},\ \bibinfo {pages}
  {176401} (\bibinfo {year} {2006})}\BibitemShut {NoStop}%
\bibitem [{\citenamefont {Kasprzak}\ \emph {et~al.}(2006)\citenamefont
  {Kasprzak}, \citenamefont {Richard}, \citenamefont {Kundermann},
  \citenamefont {Baas}, \citenamefont {Jeambrun}, \citenamefont {Keeling},
  \citenamefont {Marchetti}, \citenamefont {Szymanska}, \citenamefont {Andre},
  \citenamefont {Staehli}, \citenamefont {Savona}, \citenamefont {Littlewood},
  \citenamefont {Deveaud},\ and\ \citenamefont {Dang}}]{KASP2006}%
  \BibitemOpen
  \bibfield  {author} {\bibinfo {author} {\bibfnamefont {J.}~\bibnamefont
  {Kasprzak}}, \bibinfo {author} {\bibfnamefont {M.}~\bibnamefont {Richard}},
  \bibinfo {author} {\bibfnamefont {S.}~\bibnamefont {Kundermann}}, \bibinfo
  {author} {\bibfnamefont {A.}~\bibnamefont {Baas}}, \bibinfo {author}
  {\bibfnamefont {P.}~\bibnamefont {Jeambrun}}, \bibinfo {author}
  {\bibfnamefont {J.}~\bibnamefont {Keeling}}, \bibinfo {author} {\bibfnamefont
  {F.}~\bibnamefont {Marchetti}}, \bibinfo {author} {\bibfnamefont
  {M.}~\bibnamefont {Szymanska}}, \bibinfo {author} {\bibfnamefont
  {R.}~\bibnamefont {Andre}}, \bibinfo {author} {\bibfnamefont
  {J.}~\bibnamefont {Staehli}}, \bibinfo {author} {\bibfnamefont
  {V.}~\bibnamefont {Savona}}, \bibinfo {author} {\bibfnamefont
  {P.}~\bibnamefont {Littlewood}}, \bibinfo {author} {\bibfnamefont
  {B.}~\bibnamefont {Deveaud}}, \ and\ \bibinfo {author} {\bibfnamefont
  {L.}~\bibnamefont {Dang}},\ }\href@noop {} {\bibfield  {journal} {\bibinfo
  {journal} {Nature}\ }\textbf {\bibinfo {volume} {443}},\ \bibinfo {pages}
  {409} (\bibinfo {year} {2006})}\BibitemShut {NoStop}%
\bibitem [{\citenamefont {Klaers}\ \emph {et~al.}(2010)\citenamefont {Klaers},
  \citenamefont {Schmitt}, \citenamefont {Vewinger},\ and\ \citenamefont
  {Weitz}}]{KLAE2010}%
  \BibitemOpen
  \bibfield  {author} {\bibinfo {author} {\bibfnamefont {J.}~\bibnamefont
  {Klaers}}, \bibinfo {author} {\bibfnamefont {J.}~\bibnamefont {Schmitt}},
  \bibinfo {author} {\bibfnamefont {F.}~\bibnamefont {Vewinger}}, \ and\
  \bibinfo {author} {\bibfnamefont {M.}~\bibnamefont {Weitz}},\ }\href@noop {}
  {\bibfield  {journal} {\bibinfo  {journal} {Nature}\ }\textbf {\bibinfo
  {volume} {468}},\ \bibinfo {pages} {545} (\bibinfo {year}
  {2010})}\BibitemShut {NoStop}%
\bibitem [{\citenamefont {Schmittmann}\ and\ \citenamefont {Zia}(1995)}]{Zia}%
  \BibitemOpen
  \bibfield  {author} {\bibinfo {author} {\bibfnamefont {B.}~\bibnamefont
  {Schmittmann}}\ and\ \bibinfo {author} {\bibfnamefont {R.}~\bibnamefont
  {Zia}}\ }(\bibinfo  {publisher} {Academic Press},\ \bibinfo {year}
  {1995})\BibitemShut {NoStop}%
\bibitem [{\citenamefont {Cross}\ and\ \citenamefont
  {Hohenberg}(1993)}]{Cross}%
  \BibitemOpen
  \bibfield  {author} {\bibinfo {author} {\bibfnamefont {M.~C.}\ \bibnamefont
  {Cross}}\ and\ \bibinfo {author} {\bibfnamefont {P.~C.}\ \bibnamefont
  {Hohenberg}},\ }\href@noop {} {\bibfield  {journal} {\bibinfo  {journal}
  {Rev. Mod. Phys.}\ }\textbf {\bibinfo {volume} {65}},\ \bibinfo {pages} {851}
  (\bibinfo {year} {1993})}\BibitemShut {NoStop}%
\bibitem [{\citenamefont {Henkel}\ \emph {et~al.}(2009)\citenamefont {Henkel},
  \citenamefont {Hinrichsen},\ and\ \citenamefont {Lübeck}}]{Henkel}%
  \BibitemOpen
  \bibfield  {author} {\bibinfo {author} {\bibfnamefont {M.}~\bibnamefont
  {Henkel}}, \bibinfo {author} {\bibfnamefont {H.}~\bibnamefont {Hinrichsen}},
  \ and\ \bibinfo {author} {\bibfnamefont {S.}~\bibnamefont {Lübeck}},\
  }\href@noop {} {\emph {\bibinfo {title} {{Non-equilibrium phase
  transitions}}}},\ Theoretical and Mathematical Physics\ (\bibinfo
  {publisher} {Springer},\ \bibinfo {address} {Berlin},\ \bibinfo {year}
  {2009})\BibitemShut {NoStop}%
\bibitem [{\citenamefont {Graham}\ and\ \citenamefont {Haken}(1970)}]{Graham}%
  \BibitemOpen
  \bibfield  {author} {\bibinfo {author} {\bibfnamefont {R.}~\bibnamefont
  {Graham}}\ and\ \bibinfo {author} {\bibfnamefont {H.}~\bibnamefont {Haken}},\
  }\href@noop {} {\bibfield  {journal} {\bibinfo  {journal} {Zeitschrift für
  Physik}\ }\textbf {\bibinfo {volume} {237}},\ \bibinfo {pages} {31} (\bibinfo
  {year} {1970})}\BibitemShut {NoStop}%
\bibitem [{\citenamefont {DeGiorgio}\ and\ \citenamefont
  {Scully}(1970)}]{DEGI1970}%
  \BibitemOpen
  \bibfield  {author} {\bibinfo {author} {\bibfnamefont {V.}~\bibnamefont
  {DeGiorgio}}\ and\ \bibinfo {author} {\bibfnamefont {M.~O.}\ \bibnamefont
  {Scully}},\ }\href {\doibase 10.1103/PhysRevA.2.1170} {\bibfield  {journal}
  {\bibinfo  {journal} {Phys. Rev. A}\ }\textbf {\bibinfo {volume} {2}},\
  \bibinfo {pages} {1170} (\bibinfo {year} {1970})}\BibitemShut {NoStop}%
\bibitem [{\citenamefont {Haken}(1975)}]{HAKE1975}%
  \BibitemOpen
  \bibfield  {author} {\bibinfo {author} {\bibfnamefont {H.}~\bibnamefont
  {Haken}},\ }\href {\doibase 10.1103/RevModPhys.47.67} {\bibfield  {journal}
  {\bibinfo  {journal} {Rev. Mod. Phys.}\ }\textbf {\bibinfo {volume} {47}},\
  \bibinfo {pages} {67} (\bibinfo {year} {1975})}\BibitemShut {NoStop}%
\bibitem [{\citenamefont {Li}\ and\ \citenamefont {Iga}(2003)}]{VCSELbook}%
  \BibitemOpen
  \bibinfo {editor} {\bibfnamefont {H.~E.}\ \bibnamefont {Li}}\ and\ \bibinfo
  {editor} {\bibfnamefont {K.}~\bibnamefont {Iga}},\ eds.,\ \href@noop {}
  {\emph {\bibinfo {title} {Vertical-Cavity Surface-Emitting Laser Devices}}}\
  (\bibinfo  {publisher} {Springer Verlag},\ \bibinfo {year}
  {2003})\BibitemShut {NoStop}%
\bibitem [{\citenamefont {Chang-Hasnain}(2000)}]{VCSEL}%
  \BibitemOpen
  \bibfield  {author} {\bibinfo {author} {\bibfnamefont {C.}~\bibnamefont
  {Chang-Hasnain}},\ }\href@noop {} {\bibfield  {journal} {\bibinfo  {journal}
  {{IEEE Journal of Selected Topics in Quantum Electronics}}\ }\textbf
  {\bibinfo {volume} {6}},\ \bibinfo {pages} {978} (\bibinfo {year}
  {2000})}\BibitemShut {NoStop}%
\bibitem [{\citenamefont {{ {L. D. A. Lundeberg and G. P. Lousberg and D. L.
  Boiko and E. Kapon}}}(2007)}]{KAPON}%
  \BibitemOpen
  \bibfield  {author} {\bibinfo {author} {\bibnamefont {{ {L. D. A. Lundeberg
  and G. P. Lousberg and D. L. Boiko and E. Kapon}}}},\ }\href {\doibase
  10.1063/1.2431474} {\bibfield  {journal} {\bibinfo  {journal} {Applied
  Physics Letters}\ }\textbf {\bibinfo {volume} {90}},\ \bibinfo {eid} {021103}
  (\bibinfo {year} {2007})}\BibitemShut {NoStop}%
\bibitem [{\citenamefont {Carusotto}\ and\ \citenamefont
  {Ciuti}(2013)}]{RMP2013}%
  \BibitemOpen
  \bibfield  {author} {\bibinfo {author} {\bibfnamefont {I.}~\bibnamefont
  {Carusotto}}\ and\ \bibinfo {author} {\bibfnamefont {C.}~\bibnamefont
  {Ciuti}},\ }\href@noop {} {\bibfield  {journal} {\bibinfo  {journal} {Rev.
  Mod. Phys.}\ }\textbf {\bibinfo {volume} {85}},\ \bibinfo {pages} {299}
  (\bibinfo {year} {2013})}\BibitemShut {NoStop}%
\bibitem [{\citenamefont {Wouters}\ and\ \citenamefont
  {Carusotto}(2010)}]{WOUT2010}%
  \BibitemOpen
  \bibfield  {author} {\bibinfo {author} {\bibfnamefont {M.}~\bibnamefont
  {Wouters}}\ and\ \bibinfo {author} {\bibfnamefont {I.}~\bibnamefont
  {Carusotto}},\ }\href {\doibase 10.1103/PhysRevLett.105.020602} {\bibfield
  {journal} {\bibinfo  {journal} {Phys. Rev. Lett.}\ }\textbf {\bibinfo
  {volume} {105}},\ \bibinfo {pages} {020602} (\bibinfo {year}
  {2010})}\BibitemShut {NoStop}%
\bibitem [{\citenamefont {Richard}\ \emph {et~al.}(2005)\citenamefont
  {Richard}, \citenamefont {Kasprzak}, \citenamefont {Romestain}, \citenamefont
  {Andre},\ and\ \citenamefont {Dang}}]{Richard:PRL2005}%
  \BibitemOpen
  \bibfield  {author} {\bibinfo {author} {\bibfnamefont {M.}~\bibnamefont
  {Richard}}, \bibinfo {author} {\bibfnamefont {J.}~\bibnamefont {Kasprzak}},
  \bibinfo {author} {\bibfnamefont {R.}~\bibnamefont {Romestain}}, \bibinfo
  {author} {\bibfnamefont {R.}~\bibnamefont {Andre}}, \ and\ \bibinfo {author}
  {\bibfnamefont {L.~S.}\ \bibnamefont {Dang}},\ }\href@noop {} {\bibfield
  {journal} {\bibinfo  {journal} {Phys. Rev. Lett.}\ }\textbf {\bibinfo
  {volume} {94}},\ \bibinfo {pages} {187401} (\bibinfo {year}
  {2005})}\BibitemShut {NoStop}%
\bibitem [{\citenamefont {Wouters}\ \emph {et~al.}(2008)\citenamefont
  {Wouters}, \citenamefont {Carusotto},\ and\ \citenamefont
  {Ciuti}}]{WOUT2008}%
  \BibitemOpen
  \bibfield  {author} {\bibinfo {author} {\bibfnamefont {M.}~\bibnamefont
  {Wouters}}, \bibinfo {author} {\bibfnamefont {I.}~\bibnamefont {Carusotto}},
  \ and\ \bibinfo {author} {\bibfnamefont {C.}~\bibnamefont {Ciuti}},\ }\href
  {\doibase 10.1103/PhysRevB.77.115340} {\bibfield  {journal} {\bibinfo
  {journal} {Phys. Rev. B}\ }\textbf {\bibinfo {volume} {77}},\ \bibinfo
  {pages} {115340} (\bibinfo {year} {2008})}\BibitemShut {NoStop}%
\bibitem [{\citenamefont {Wertz}\ \emph {et~al.}(2010)\citenamefont {Wertz},
  \citenamefont {Ferrier}, \citenamefont {Solnyshkov}, \citenamefont {Johne},
  \citenamefont {Sanvitto}, \citenamefont {Lemaître}, \citenamefont {Sagnes},
  \citenamefont {Grousson}, \citenamefont {Kavokin}, \citenamefont {Senellart},
  \citenamefont {Malpuech},\ and\ \citenamefont {Bloch}}]{WERTZ}%
  \BibitemOpen
  \bibfield  {author} {\bibinfo {author} {\bibfnamefont {E.}~\bibnamefont
  {Wertz}}, \bibinfo {author} {\bibfnamefont {L.}~\bibnamefont {Ferrier}},
  \bibinfo {author} {\bibfnamefont {D.}~\bibnamefont {Solnyshkov}}, \bibinfo
  {author} {\bibfnamefont {R.}~\bibnamefont {Johne}}, \bibinfo {author}
  {\bibfnamefont {D.}~\bibnamefont {Sanvitto}}, \bibinfo {author}
  {\bibfnamefont {A.}~\bibnamefont {Lemaître}}, \bibinfo {author}
  {\bibfnamefont {I.}~\bibnamefont {Sagnes}}, \bibinfo {author} {\bibfnamefont
  {R.}~\bibnamefont {Grousson}}, \bibinfo {author} {\bibfnamefont
  {A.}~\bibnamefont {Kavokin}}, \bibinfo {author} {\bibfnamefont
  {P.}~\bibnamefont {Senellart}}, \bibinfo {author} {\bibfnamefont
  {G.}~\bibnamefont {Malpuech}}, \ and\ \bibinfo {author} {\bibfnamefont
  {J.}~\bibnamefont {Bloch}},\ }\href@noop {} {\bibfield  {journal} {\bibinfo
  {journal} {Nat. Phys.}\ }\textbf {\bibinfo {volume} {6}},\ \bibinfo {pages}
  {860} (\bibinfo {year} {2010})}\BibitemShut {NoStop}%
\bibitem [{\citenamefont {Bajoni}\ \emph {et~al.}(2007)\citenamefont {Bajoni},
  \citenamefont {Senellart}, \citenamefont {Lemaitre},\ and\ \citenamefont
  {Bloch}}]{BAJO2007}%
  \BibitemOpen
  \bibfield  {author} {\bibinfo {author} {\bibfnamefont {D.}~\bibnamefont
  {Bajoni}}, \bibinfo {author} {\bibfnamefont {P.}~\bibnamefont {Senellart}},
  \bibinfo {author} {\bibfnamefont {A.}~\bibnamefont {Lemaitre}}, \ and\
  \bibinfo {author} {\bibfnamefont {J.}~\bibnamefont {Bloch}},\ }\href
  {\doibase 10.1103/PhysRevB.76.201305} {\bibfield  {journal} {\bibinfo
  {journal} {Phys. Rev. B}\ }\textbf {\bibinfo {volume} {76}},\ \bibinfo
  {pages} {201305} (\bibinfo {year} {2007})}\BibitemShut {NoStop}%
\bibitem [{\citenamefont {Kirton}\ and\ \citenamefont
  {Keeling}(2013)}]{KEELING2013}%
  \BibitemOpen
  \bibfield  {author} {\bibinfo {author} {\bibfnamefont {P.}~\bibnamefont
  {Kirton}}\ and\ \bibinfo {author} {\bibfnamefont {J.}~\bibnamefont
  {Keeling}},\ }\href@noop {} {\bibfield  {journal} {\bibinfo  {journal}
  {Physical Review Letters}\ }\textbf {\bibinfo {volume} {111}},\ \bibinfo
  {pages} {100404} (\bibinfo {year} {2013})}\BibitemShut {NoStop}%
\bibitem [{\citenamefont {{Szymanska}}\ \emph {et~al.}(2006)\citenamefont
  {{Szymanska}}, \citenamefont {Keeling},\ and\ \citenamefont
  {Littlewood}}]{SZYM2006}%
  \BibitemOpen
  \bibfield  {author} {\bibinfo {author} {\bibfnamefont {M.~H.}\ \bibnamefont
  {{Szymanska}}}, \bibinfo {author} {\bibfnamefont {J.}~\bibnamefont
  {Keeling}}, \ and\ \bibinfo {author} {\bibfnamefont {P.~B.}\ \bibnamefont
  {Littlewood}},\ }\href {\doibase 10.1103/PhysRevLett.96.230602} {\bibfield
  {journal} {\bibinfo  {journal} {Phys. Rev. Lett.}\ }\textbf {\bibinfo
  {volume} {96}},\ \bibinfo {pages} {230602} (\bibinfo {year}
  {2006})}\BibitemShut {NoStop}%
\bibitem [{\citenamefont {Wouters}\ and\ \citenamefont
  {Carusotto}(2006)}]{WOUT2006}%
  \BibitemOpen
  \bibfield  {author} {\bibinfo {author} {\bibfnamefont {M.}~\bibnamefont
  {Wouters}}\ and\ \bibinfo {author} {\bibfnamefont {I.}~\bibnamefont
  {Carusotto}},\ }\href {\doibase 10.1103/PhysRevB.74.245316} {\bibfield
  {journal} {\bibinfo  {journal} {Phys. Rev. B}\ }\textbf {\bibinfo {volume}
  {74}},\ \bibinfo {pages} {245316} (\bibinfo {year} {2006})}\BibitemShut
  {NoStop}%
\bibitem [{\citenamefont {Wouters}\ and\ \citenamefont
  {Carusotto}(2007{\natexlab{a}})}]{WOUT2007}%
  \BibitemOpen
  \bibfield  {author} {\bibinfo {author} {\bibfnamefont {M.}~\bibnamefont
  {Wouters}}\ and\ \bibinfo {author} {\bibfnamefont {I.}~\bibnamefont
  {Carusotto}},\ }\href {\doibase 10.1103/PhysRevLett.99.140402} {\bibfield
  {journal} {\bibinfo  {journal} {Phys. Rev. Lett.}\ }\textbf {\bibinfo
  {volume} {99}},\ \bibinfo {pages} {140402} (\bibinfo {year}
  {2007}{\natexlab{a}})}\BibitemShut {NoStop}%
\bibitem [{\citenamefont {Sieberer}\ \emph {et~al.}(2013)\citenamefont
  {Sieberer}, \citenamefont {Huber}, \citenamefont {Altman},\ and\
  \citenamefont {Diehl}}]{Sieberer2013}%
  \BibitemOpen
  \bibfield  {author} {\bibinfo {author} {\bibfnamefont {L.~M.}\ \bibnamefont
  {Sieberer}}, \bibinfo {author} {\bibfnamefont {S.~D.}\ \bibnamefont {Huber}},
  \bibinfo {author} {\bibfnamefont {E.}~\bibnamefont {Altman}}, \ and\ \bibinfo
  {author} {\bibfnamefont {S.}~\bibnamefont {Diehl}},\ }\href@noop {}
  {\bibfield  {journal} {\bibinfo  {journal} {Phys. Rev. Lett.}\ }\textbf
  {\bibinfo {volume} {110}},\ \bibinfo {pages} {195301} (\bibinfo {year}
  {2013})}\BibitemShut {NoStop}%
\bibitem [{\citenamefont {Fischer}\ and\ \citenamefont
  {Weill}(2012)}]{Fischer}%
  \BibitemOpen
  \bibfield  {author} {\bibinfo {author} {\bibfnamefont {B.}~\bibnamefont
  {Fischer}}\ and\ \bibinfo {author} {\bibfnamefont {R.}~\bibnamefont
  {Weill}},\ }\href@noop {} {\bibfield  {journal} {\bibinfo  {journal} {Opt.
  Express}\ }\textbf {\bibinfo {volume} {20}},\ \bibinfo {pages} {26704}
  (\bibinfo {year} {2012})}\BibitemShut {NoStop}%
\bibitem [{\citenamefont {Hafezi}\ \emph {et~al.}(2014)\citenamefont {Hafezi},
  \citenamefont {Adhikari},\ and\ \citenamefont {Taylor}}]{hafezi2014chemical}%
  \BibitemOpen
  \bibfield  {author} {\bibinfo {author} {\bibfnamefont {M.}~\bibnamefont
  {Hafezi}}, \bibinfo {author} {\bibfnamefont {P.}~\bibnamefont {Adhikari}}, \
  and\ \bibinfo {author} {\bibfnamefont {J.}~\bibnamefont {Taylor}},\
  }\href@noop {} {\bibfield  {journal} {\bibinfo  {journal} {arXiv preprint
  arXiv:1405.5821}\ } (\bibinfo {year} {2014})}\BibitemShut {NoStop}%
\bibitem [{Note1()}]{Note1}%
  \BibitemOpen
  \bibinfo {note} {{Of course, BEC remains possible if a harmonic trap
  potential is added to the 2D gas~\cite {PITA2004}, as done in the experiment
  of~\cite {KLAE2010}}}\BibitemShut {NoStop}%
\bibitem [{\citenamefont {Mermin}\ and\ \citenamefont
  {Wagner}(1966)}]{MERM1966}%
  \BibitemOpen
  \bibfield  {author} {\bibinfo {author} {\bibfnamefont {N.~D.}\ \bibnamefont
  {Mermin}}\ and\ \bibinfo {author} {\bibfnamefont {H.}~\bibnamefont
  {Wagner}},\ }\href {\doibase 10.1103/PhysRevLett.17.1133} {\bibfield
  {journal} {\bibinfo  {journal} {Phys. Rev. Lett.}\ }\textbf {\bibinfo
  {volume} {17}},\ \bibinfo {pages} {1133} (\bibinfo {year}
  {1966})}\BibitemShut {NoStop}%
\bibitem [{\citenamefont {Chiocchetta}\ and\ \citenamefont
  {Carusotto}(2013)}]{ACIC2013}%
  \BibitemOpen
  \bibfield  {author} {\bibinfo {author} {\bibfnamefont {A.}~\bibnamefont
  {Chiocchetta}}\ and\ \bibinfo {author} {\bibfnamefont {I.}~\bibnamefont
  {Carusotto}},\ }\href {http://stacks.iop.org/0295-5075/102/i=6/a=67007}
  {\bibfield  {journal} {\bibinfo  {journal} {EPL (Europhysics Letters)}\
  }\textbf {\bibinfo {volume} {102}},\ \bibinfo {pages} {67007} (\bibinfo
  {year} {2013})}\BibitemShut {NoStop}%
\bibitem [{\citenamefont {Spano}\ \emph {et~al.}(2012)\citenamefont {Spano},
  \citenamefont {Cuadra}, \citenamefont {Tosi}, \citenamefont {Ant{\'o}n},
  \citenamefont {Lingg}, \citenamefont {Sanvitto}, \citenamefont {Mart{\'i}n},
  \citenamefont {Vi{\~n}a}, \citenamefont {Eastham}, \citenamefont {{van der
  Poel}},\ and\ \citenamefont {Hvam}}]{SPAN2012}%
  \BibitemOpen
  \bibfield  {author} {\bibinfo {author} {\bibfnamefont {R.}~\bibnamefont
  {Spano}}, \bibinfo {author} {\bibfnamefont {J.}~\bibnamefont {Cuadra}},
  \bibinfo {author} {\bibfnamefont {G.}~\bibnamefont {Tosi}}, \bibinfo {author}
  {\bibfnamefont {C.}~\bibnamefont {Ant{\'o}n}}, \bibinfo {author}
  {\bibfnamefont {C.~A.}\ \bibnamefont {Lingg}}, \bibinfo {author}
  {\bibfnamefont {D.}~\bibnamefont {Sanvitto}}, \bibinfo {author}
  {\bibfnamefont {M.~D.}\ \bibnamefont {Mart{\'i}n}}, \bibinfo {author}
  {\bibfnamefont {L.}~\bibnamefont {Vi{\~n}a}}, \bibinfo {author}
  {\bibfnamefont {P.~R.}\ \bibnamefont {Eastham}}, \bibinfo {author}
  {\bibfnamefont {M.}~\bibnamefont {{van der Poel}}}, \ and\ \bibinfo {author}
  {\bibfnamefont {J.~M.}\ \bibnamefont {Hvam}},\ }\href@noop {} {\bibfield
  {journal} {\bibinfo  {journal} {New Journal of Physics}\ }\textbf {\bibinfo
  {volume} {14}},\ \bibinfo {pages} {075018} (\bibinfo {year}
  {2012})}\BibitemShut {NoStop}%
\bibitem [{\citenamefont {Roumpos}\ \emph {et~al.}(2012)\citenamefont
  {Roumpos}, \citenamefont {Lohse}, \citenamefont {Nitsche}, \citenamefont
  {Keeling}, \citenamefont {Szyma{\'n}ska}, \citenamefont {Littlewood},
  \citenamefont {L{\"o}ffler}, \citenamefont {H{\"o}fling}, \citenamefont
  {Worschech}, \citenamefont {Forchel},\ and\ \citenamefont
  {Yamamoto}}]{ROUM2012}%
  \BibitemOpen
  \bibfield  {author} {\bibinfo {author} {\bibfnamefont {G.}~\bibnamefont
  {Roumpos}}, \bibinfo {author} {\bibfnamefont {M.}~\bibnamefont {Lohse}},
  \bibinfo {author} {\bibfnamefont {W.~H.}\ \bibnamefont {Nitsche}}, \bibinfo
  {author} {\bibfnamefont {J.}~\bibnamefont {Keeling}}, \bibinfo {author}
  {\bibfnamefont {M.~H.}\ \bibnamefont {Szyma{\'n}ska}}, \bibinfo {author}
  {\bibfnamefont {P.~B.}\ \bibnamefont {Littlewood}}, \bibinfo {author}
  {\bibfnamefont {A.}~\bibnamefont {L{\"o}ffler}}, \bibinfo {author}
  {\bibfnamefont {S.}~\bibnamefont {H{\"o}fling}}, \bibinfo {author}
  {\bibfnamefont {L.}~\bibnamefont {Worschech}}, \bibinfo {author}
  {\bibfnamefont {A.}~\bibnamefont {Forchel}}, \ and\ \bibinfo {author}
  {\bibfnamefont {Y.}~\bibnamefont {Yamamoto}},\ }\href@noop {} {\bibfield
  {journal} {\bibinfo  {journal} {Proceedings of the National Academy of
  Sciences}\ } (\bibinfo {year} {2012})}\BibitemShut {NoStop}%
\bibitem [{\citenamefont {Altman}\ \emph {et~al.}(2013)\citenamefont {Altman},
  \citenamefont {Sieberer}, \citenamefont {Chen}, \citenamefont {Diehl},\ and\
  \citenamefont {Toner}}]{Altman}%
  \BibitemOpen
  \bibfield  {author} {\bibinfo {author} {\bibfnamefont {E.}~\bibnamefont
  {Altman}}, \bibinfo {author} {\bibfnamefont {L.~M.}\ \bibnamefont
  {Sieberer}}, \bibinfo {author} {\bibfnamefont {L.}~\bibnamefont {Chen}},
  \bibinfo {author} {\bibfnamefont {S.}~\bibnamefont {Diehl}}, \ and\ \bibinfo
  {author} {\bibfnamefont {J.}~\bibnamefont {Toner}},\ }\href@noop {}
  {\bibfield  {journal} {\bibinfo  {journal} {ArXiv e-prints}\ } (\bibinfo
  {year} {2013})},\ \Eprint {http://arxiv.org/abs/1311.0876} {1311.0876
  [cond-mat.stat-mech]} \BibitemShut {NoStop}%
\bibitem [{\citenamefont {Gladilin}\ \emph {et~al.}(2013)\citenamefont
  {Gladilin}, \citenamefont {Ji},\ and\ \citenamefont {Wouters}}]{Wouters2013}%
  \BibitemOpen
  \bibfield  {author} {\bibinfo {author} {\bibfnamefont {V.}~\bibnamefont
  {Gladilin}}, \bibinfo {author} {\bibfnamefont {K.}~\bibnamefont {Ji}}, \ and\
  \bibinfo {author} {\bibfnamefont {M.}~\bibnamefont {Wouters}},\ }\href@noop
  {} {\bibfield  {journal} {\bibinfo  {journal} {ArXiv e-prints}\ } (\bibinfo
  {year} {2013})},\ \Eprint {http://arxiv.org/abs/1312.0452} {1312.0452
  [cond-mat.quant-gas]} \BibitemShut {NoStop}%
\bibitem [{\citenamefont {Carusotto}\ and\ \citenamefont
  {Ciuti}(2005)}]{Ciuti2005}%
  \BibitemOpen
  \bibfield  {author} {\bibinfo {author} {\bibfnamefont {I.}~\bibnamefont
  {Carusotto}}\ and\ \bibinfo {author} {\bibfnamefont {C.}~\bibnamefont
  {Ciuti}},\ }\href@noop {} {\bibfield  {journal} {\bibinfo  {journal} {Phys.
  Rev. B}\ }\textbf {\bibinfo {volume} {72}},\ \bibinfo {pages} {125335}
  (\bibinfo {year} {2005})}\BibitemShut {NoStop}%
\bibitem [{\citenamefont {Wouters}\ and\ \citenamefont
  {Savona}(2009)}]{WOUT2009}%
  \BibitemOpen
  \bibfield  {author} {\bibinfo {author} {\bibfnamefont {M.}~\bibnamefont
  {Wouters}}\ and\ \bibinfo {author} {\bibfnamefont {V.}~\bibnamefont
  {Savona}},\ }\href {\doibase 10.1103/PhysRevB.79.165302} {\bibfield
  {journal} {\bibinfo  {journal} {Phys. Rev. B}\ }\textbf {\bibinfo {volume}
  {79}},\ \bibinfo {pages} {165302} (\bibinfo {year} {2009})}\BibitemShut
  {NoStop}%
\bibitem [{\citenamefont {Keeling}\ and\ \citenamefont
  {Berloff}(2008)}]{Keeling}%
  \BibitemOpen
  \bibfield  {author} {\bibinfo {author} {\bibfnamefont {J.}~\bibnamefont
  {Keeling}}\ and\ \bibinfo {author} {\bibfnamefont {N.~G.}\ \bibnamefont
  {Berloff}},\ }\href@noop {} {\bibfield  {journal} {\bibinfo  {journal} {Phys.
  Rev. Lett.}\ }\textbf {\bibinfo {volume} {100}},\ \bibinfo {pages} {250401}
  (\bibinfo {year} {2008})}\BibitemShut {NoStop}%
\bibitem [{\citenamefont {Gardiner}\ and\ \citenamefont
  {Zoller}(2000)}]{GARD2000}%
  \BibitemOpen
  \bibfield  {author} {\bibinfo {author} {\bibfnamefont {C.}~\bibnamefont
  {Gardiner}}\ and\ \bibinfo {author} {\bibfnamefont {P.}~\bibnamefont
  {Zoller}},\ }\href@noop {} {\emph {\bibinfo {title} {Quantum Noise}}},\
  \bibinfo {edition} {2nd}\ ed.\ (\bibinfo  {publisher} {Springer-Verlag},\
  \bibinfo {year} {2000})\BibitemShut {NoStop}%
\bibitem [{\citenamefont {Lax}(1966)}]{LAX1966}%
  \BibitemOpen
  \bibfield  {author} {\bibinfo {author} {\bibfnamefont {M.}~\bibnamefont
  {Lax}},\ }\href {\doibase 10.1103/PhysRev.145.110} {\bibfield  {journal}
  {\bibinfo  {journal} {Phys. Rev.}\ }\textbf {\bibinfo {volume} {145}},\
  \bibinfo {pages} {110} (\bibinfo {year} {1966})}\BibitemShut {NoStop}%
\bibitem [{\citenamefont {Louisell}(1973)}]{LOUI1973}%
  \BibitemOpen
  \bibfield  {author} {\bibinfo {author} {\bibfnamefont {W.}~\bibnamefont
  {Louisell}},\ }\href@noop {} {\emph {\bibinfo {title} {Quantum Statistical
  Properties of Radiation}}}\ (\bibinfo  {publisher} {Wiley, New York},\
  \bibinfo {year} {1973})\BibitemShut {NoStop}%
\bibitem [{\citenamefont {Gordon}(1967)}]{GORD1967}%
  \BibitemOpen
  \bibfield  {author} {\bibinfo {author} {\bibfnamefont {J.~P.}\ \bibnamefont
  {Gordon}},\ }\href {\doibase 10.1103/PhysRev.161.367} {\bibfield  {journal}
  {\bibinfo  {journal} {Phys. Rev.}\ }\textbf {\bibinfo {volume} {161}},\
  \bibinfo {pages} {367} (\bibinfo {year} {1967})}\BibitemShut {NoStop}%
\bibitem [{\citenamefont {Haken}(1984)}]{HAKE1984}%
  \BibitemOpen
  \bibfield  {author} {\bibinfo {author} {\bibfnamefont {H.}~\bibnamefont
  {Haken}},\ }\href@noop {} {\emph {\bibinfo {title} {Laser Theory}}}\
  (\bibinfo  {publisher} {Springer-Verlag, Berlin},\ \bibinfo {year}
  {1984})\BibitemShut {NoStop}%
\bibitem [{\citenamefont {Keeling}\ \emph {et~al.}(2007)\citenamefont
  {Keeling}, \citenamefont {Marchetti}, \citenamefont {Szyma{\'n}ska},\ and\
  \citenamefont {Littlewood}}]{keeling2007}%
  \BibitemOpen
  \bibfield  {author} {\bibinfo {author} {\bibfnamefont {J.}~\bibnamefont
  {Keeling}}, \bibinfo {author} {\bibfnamefont {F.}~\bibnamefont {Marchetti}},
  \bibinfo {author} {\bibfnamefont {M.}~\bibnamefont {Szyma{\'n}ska}}, \ and\
  \bibinfo {author} {\bibfnamefont {P.}~\bibnamefont {Littlewood}},\
  }\href@noop {} {\bibfield  {journal} {\bibinfo  {journal} {Semiconductor
  science and technology}\ }\textbf {\bibinfo {volume} {22}},\ \bibinfo {pages}
  {R1} (\bibinfo {year} {2007})}\BibitemShut {NoStop}%
\bibitem [{\citenamefont {Deng}\ \emph {et~al.}(2010)\citenamefont {Deng},
  \citenamefont {Haug},\ and\ \citenamefont {Yamamoto}}]{DENG2010}%
  \BibitemOpen
  \bibfield  {author} {\bibinfo {author} {\bibfnamefont {H.}~\bibnamefont
  {Deng}}, \bibinfo {author} {\bibfnamefont {H.}~\bibnamefont {Haug}}, \ and\
  \bibinfo {author} {\bibfnamefont {Y.}~\bibnamefont {Yamamoto}},\ }\href
  {\doibase 10.1103/RevModPhys.82.1489} {\bibfield  {journal} {\bibinfo
  {journal} {Rev. Mod. Phys.}\ }\textbf {\bibinfo {volume} {82}},\ \bibinfo
  {pages} {1489} (\bibinfo {year} {2010})}\BibitemShut {NoStop}%
\bibitem [{\citenamefont {de~Leeuw}\ \emph {et~al.}(2013)\citenamefont
  {de~Leeuw}, \citenamefont {Stoof},\ and\ \citenamefont
  {Duine}}]{DELEEUW2013}%
  \BibitemOpen
  \bibfield  {author} {\bibinfo {author} {\bibfnamefont {A.-W.}\ \bibnamefont
  {de~Leeuw}}, \bibinfo {author} {\bibfnamefont {H.~T.~C.}\ \bibnamefont
  {Stoof}}, \ and\ \bibinfo {author} {\bibfnamefont {R.~A.}\ \bibnamefont
  {Duine}},\ }\href@noop {} {\bibfield  {journal} {\bibinfo  {journal} {Phys.
  Rev. A}\ }\textbf {\bibinfo {volume} {88}},\ \bibinfo {pages} {033829}
  (\bibinfo {year} {2013})}\BibitemShut {NoStop}%
\bibitem [{\citenamefont {Ciuti}\ and\ \citenamefont
  {Carusotto}(2006)}]{ICCCPRA2006}%
  \BibitemOpen
  \bibfield  {author} {\bibinfo {author} {\bibfnamefont {C.}~\bibnamefont
  {Ciuti}}\ and\ \bibinfo {author} {\bibfnamefont {I.}~\bibnamefont
  {Carusotto}},\ }\href {\doibase 10.1103/PhysRevA.74.033811} {\bibfield
  {journal} {\bibinfo  {journal} {Phys. Rev. A}\ }\textbf {\bibinfo {volume}
  {74}},\ \bibinfo {pages} {033811} (\bibinfo {year} {2006})}\BibitemShut
  {NoStop}%
\bibitem [{\citenamefont {Cohen-Tannoudji}\ \emph {et~al.}(2004)\citenamefont
  {Cohen-Tannoudji}, \citenamefont {Dupont-Roc},\ and\ \citenamefont
  {Grynberg}}]{COHE2004}%
  \BibitemOpen
  \bibfield  {author} {\bibinfo {author} {\bibfnamefont {C.}~\bibnamefont
  {Cohen-Tannoudji}}, \bibinfo {author} {\bibfnamefont {J.}~\bibnamefont
  {Dupont-Roc}}, \ and\ \bibinfo {author} {\bibfnamefont {G.}~\bibnamefont
  {Grynberg}},\ }\href@noop {} {\emph {\bibinfo {title} {Atom-Photon
  Interactions: Basic Processes and Applications}}}\ (\bibinfo  {publisher}
  {Wiley-VCH Verlag GmbH \& Co. KGaA, Weinheim},\ \bibinfo {year}
  {2004})\BibitemShut {NoStop}%
\bibitem [{\citenamefont {Scully}\ and\ \citenamefont
  {Zubairy}(1997)}]{SCUL1997}%
  \BibitemOpen
  \bibfield  {author} {\bibinfo {author} {\bibfnamefont {M.}~\bibnamefont
  {Scully}}\ and\ \bibinfo {author} {\bibfnamefont {M.}~\bibnamefont
  {Zubairy}},\ }\href@noop {} {\emph {\bibinfo {title} {Quantum Optics}}}\
  (\bibinfo  {publisher} {Cambridge University Press},\ \bibinfo {year}
  {1997})\BibitemShut {NoStop}%
\bibitem [{\citenamefont {Wouters}\ and\ \citenamefont
  {Carusotto}(2007{\natexlab{b}})}]{Wouters2007}%
  \BibitemOpen
  \bibfield  {author} {\bibinfo {author} {\bibfnamefont {M.}~\bibnamefont
  {Wouters}}\ and\ \bibinfo {author} {\bibfnamefont {I.}~\bibnamefont
  {Carusotto}},\ }\href@noop {} {\bibfield  {journal} {\bibinfo  {journal}
  {Phys. Rev. B}\ }\textbf {\bibinfo {volume} {75}},\ \bibinfo {pages} {075332}
  (\bibinfo {year} {2007}{\natexlab{b}})}\BibitemShut {NoStop}%
\bibitem [{\citenamefont {Christmann}\ \emph {et~al.}(2012)\citenamefont
  {Christmann}, \citenamefont {Tosi}, \citenamefont {Berloff}, \citenamefont
  {Tsotsis}, \citenamefont {Eldridge}, \citenamefont {Hatzopoulos},
  \citenamefont {Savvidis},\ and\ \citenamefont {Baumberg}}]{BAUM}%
  \BibitemOpen
  \bibfield  {author} {\bibinfo {author} {\bibfnamefont {G.}~\bibnamefont
  {Christmann}}, \bibinfo {author} {\bibfnamefont {G.}~\bibnamefont {Tosi}},
  \bibinfo {author} {\bibfnamefont {N.~G.}\ \bibnamefont {Berloff}}, \bibinfo
  {author} {\bibfnamefont {P.}~\bibnamefont {Tsotsis}}, \bibinfo {author}
  {\bibfnamefont {P.~S.}\ \bibnamefont {Eldridge}}, \bibinfo {author}
  {\bibfnamefont {Z.}~\bibnamefont {Hatzopoulos}}, \bibinfo {author}
  {\bibfnamefont {P.~G.}\ \bibnamefont {Savvidis}}, \ and\ \bibinfo {author}
  {\bibfnamefont {J.~J.}\ \bibnamefont {Baumberg}},\ }\href {\doibase
  10.1103/PhysRevB.85.235303} {\bibfield  {journal} {\bibinfo  {journal} {Phys.
  Rev. B}\ }\textbf {\bibinfo {volume} {85}},\ \bibinfo {pages} {235303}
  (\bibinfo {year} {2012})}\BibitemShut {NoStop}%
\bibitem [{\citenamefont {Castin}\ and\ \citenamefont {Dum}(1998)}]{castindum}%
  \BibitemOpen
  \bibfield  {author} {\bibinfo {author} {\bibfnamefont {Y.}~\bibnamefont
  {Castin}}\ and\ \bibinfo {author} {\bibfnamefont {R.}~\bibnamefont {Dum}},\
  }\href {\doibase 10.1103/PhysRevA.57.3008} {\bibfield  {journal} {\bibinfo
  {journal} {Phys. Rev. A}\ }\textbf {\bibinfo {volume} {57}},\ \bibinfo
  {pages} {3008} (\bibinfo {year} {1998})}\BibitemShut {NoStop}%
\bibitem [{\citenamefont {Sarchi}\ and\ \citenamefont
  {Carusotto}(2010)}]{Sarchi:2010PRB}%
  \BibitemOpen
  \bibfield  {author} {\bibinfo {author} {\bibfnamefont {D.}~\bibnamefont
  {Sarchi}}\ and\ \bibinfo {author} {\bibfnamefont {I.}~\bibnamefont
  {Carusotto}},\ }\href {\doibase 10.1103/PhysRevB.81.075320} {\bibfield
  {journal} {\bibinfo  {journal} {Phys. Rev. B}\ }\textbf {\bibinfo {volume}
  {81}},\ \bibinfo {pages} {075320} (\bibinfo {year} {2010})}\BibitemShut
  {NoStop}%
\bibitem [{\citenamefont {Le~Boit\'e}\ \emph {et~al.}(2013)\citenamefont
  {Le~Boit\'e}, \citenamefont {Orso},\ and\ \citenamefont
  {Ciuti}}]{supersolid2}%
  \BibitemOpen
  \bibfield  {author} {\bibinfo {author} {\bibfnamefont {A.}~\bibnamefont
  {Le~Boit\'e}}, \bibinfo {author} {\bibfnamefont {G.}~\bibnamefont {Orso}}, \
  and\ \bibinfo {author} {\bibfnamefont {C.}~\bibnamefont {Ciuti}},\ }\href
  {\doibase 10.1103/PhysRevLett.110.233601} {\bibfield  {journal} {\bibinfo
  {journal} {Phys. Rev. Lett.}\ }\textbf {\bibinfo {volume} {110}},\ \bibinfo
  {pages} {233601} (\bibinfo {year} {2013})}\BibitemShut {NoStop}%
\bibitem [{\citenamefont {Jin}\ \emph {et~al.}(2013)\citenamefont {Jin},
  \citenamefont {Rossini}, \citenamefont {Fazio}, \citenamefont {Leib},\ and\
  \citenamefont {Hartmann}}]{supersolid1}%
  \BibitemOpen
  \bibfield  {author} {\bibinfo {author} {\bibfnamefont {J.}~\bibnamefont
  {Jin}}, \bibinfo {author} {\bibfnamefont {D.}~\bibnamefont {Rossini}},
  \bibinfo {author} {\bibfnamefont {R.}~\bibnamefont {Fazio}}, \bibinfo
  {author} {\bibfnamefont {M.}~\bibnamefont {Leib}}, \ and\ \bibinfo {author}
  {\bibfnamefont {M.~J.}\ \bibnamefont {Hartmann}},\ }\href {\doibase
  10.1103/PhysRevLett.110.163605} {\bibfield  {journal} {\bibinfo  {journal}
  {Phys. Rev. Lett.}\ }\textbf {\bibinfo {volume} {110}},\ \bibinfo {pages}
  {163605} (\bibinfo {year} {2013})}\BibitemShut {NoStop}%
\bibitem [{\citenamefont {Gardiner}(2004)}]{GARD2004}%
  \BibitemOpen
  \bibfield  {author} {\bibinfo {author} {\bibfnamefont {C.}~\bibnamefont
  {Gardiner}},\ }\href@noop {} {\emph {\bibinfo {title} {Handbook of stochastic
  methods for physics, chemistry and the natural sciences}}},\ \bibinfo
  {edition} {3rd}\ ed.,\ Springer Series in Synergetics\ (\bibinfo  {publisher}
  {Springer-Verlag},\ \bibinfo {year} {2004})\BibitemShut {NoStop}%
\bibitem [{\citenamefont {Marchetti}\ \emph {et~al.}(2007)\citenamefont
  {Marchetti}, \citenamefont {Keeling}, \citenamefont
  {Szyma\ifmmode~\acute{n}\else \'{n}\fi{}ska},\ and\ \citenamefont
  {Littlewood}}]{Marchetti2007}%
  \BibitemOpen
  \bibfield  {author} {\bibinfo {author} {\bibfnamefont {F.~M.}\ \bibnamefont
  {Marchetti}}, \bibinfo {author} {\bibfnamefont {J.}~\bibnamefont {Keeling}},
  \bibinfo {author} {\bibfnamefont {M.~H.}\ \bibnamefont
  {Szyma\ifmmode~\acute{n}\else \'{n}\fi{}ska}}, \ and\ \bibinfo {author}
  {\bibfnamefont {P.~B.}\ \bibnamefont {Littlewood}},\ }\href {\doibase
  10.1103/PhysRevB.76.115326} {\bibfield  {journal} {\bibinfo  {journal} {Phys.
  Rev. B}\ }\textbf {\bibinfo {volume} {76}},\ \bibinfo {pages} {115326}
  (\bibinfo {year} {2007})}\BibitemShut {NoStop}%
\bibitem [{\citenamefont {Byrnes}\ \emph {et~al.}(2012)\citenamefont {Byrnes},
  \citenamefont {Horikiri}, \citenamefont {Ishida}, \citenamefont {Fraser},\
  and\ \citenamefont {Yamamoto}}]{Byrnes}%
  \BibitemOpen
  \bibfield  {author} {\bibinfo {author} {\bibfnamefont {T.}~\bibnamefont
  {Byrnes}}, \bibinfo {author} {\bibfnamefont {T.}~\bibnamefont {Horikiri}},
  \bibinfo {author} {\bibfnamefont {N.}~\bibnamefont {Ishida}}, \bibinfo
  {author} {\bibfnamefont {M.}~\bibnamefont {Fraser}}, \ and\ \bibinfo {author}
  {\bibfnamefont {Y.}~\bibnamefont {Yamamoto}},\ }\href@noop {} {\bibfield
  {journal} {\bibinfo  {journal} {Phys. Rev. B}\ }\textbf {\bibinfo {volume}
  {85}},\ \bibinfo {pages} {075130} (\bibinfo {year} {2012})}\BibitemShut
  {NoStop}%
\bibitem [{\citenamefont {Lugiato}\ \emph {et~al.}(1984)\citenamefont
  {Lugiato}, \citenamefont {Mandel},\ and\ \citenamefont
  {Narducci}}]{LUGI1984}%
  \BibitemOpen
  \bibfield  {author} {\bibinfo {author} {\bibfnamefont {L.~A.}\ \bibnamefont
  {Lugiato}}, \bibinfo {author} {\bibfnamefont {P.}~\bibnamefont {Mandel}}, \
  and\ \bibinfo {author} {\bibfnamefont {L.~M.}\ \bibnamefont {Narducci}},\
  }\href {\doibase 10.1103/PhysRevA.29.1438} {\bibfield  {journal} {\bibinfo
  {journal} {Phys. Rev. A}\ }\textbf {\bibinfo {volume} {29}},\ \bibinfo
  {pages} {1438} (\bibinfo {year} {1984})}\BibitemShut {NoStop}%
\bibitem [{\citenamefont {Lax}\ and\ \citenamefont {Louisell}(1967)}]{LAX_IX}%
  \BibitemOpen
  \bibfield  {author} {\bibinfo {author} {\bibfnamefont {M.}~\bibnamefont
  {Lax}}\ and\ \bibinfo {author} {\bibfnamefont {W.}~\bibnamefont {Louisell}},\
  }\href@noop {} {\bibfield  {journal} {\bibinfo  {journal} {Quantum
  Electronics, IEEE Journal of}\ }\textbf {\bibinfo {volume} {3}},\ \bibinfo
  {pages} {47} (\bibinfo {year} {1967})}\BibitemShut {NoStop}%
\bibitem [{\citenamefont {Lax}\ and\ \citenamefont {Yuen}(1968)}]{LAX1968}%
  \BibitemOpen
  \bibfield  {author} {\bibinfo {author} {\bibfnamefont {M.}~\bibnamefont
  {Lax}}\ and\ \bibinfo {author} {\bibfnamefont {H.}~\bibnamefont {Yuen}},\
  }\href@noop {} {\bibfield  {journal} {\bibinfo  {journal} {Phys. Rev.}\
  }\textbf {\bibinfo {volume} {172}},\ \bibinfo {pages} {362} (\bibinfo {year}
  {1968})}\BibitemShut {NoStop}%
\bibitem [{\citenamefont {Benkert}\ \emph {et~al.}(1990)\citenamefont
  {Benkert}, \citenamefont {Scully}, \citenamefont {Bergou}, \citenamefont
  {Davidovich}, \citenamefont {Hillery},\ and\ \citenamefont
  {Orszag}}]{Benkert}%
  \BibitemOpen
  \bibfield  {author} {\bibinfo {author} {\bibfnamefont {C.}~\bibnamefont
  {Benkert}}, \bibinfo {author} {\bibfnamefont {M.~O.}\ \bibnamefont {Scully}},
  \bibinfo {author} {\bibfnamefont {J.}~\bibnamefont {Bergou}}, \bibinfo
  {author} {\bibfnamefont {L.}~\bibnamefont {Davidovich}}, \bibinfo {author}
  {\bibfnamefont {M.}~\bibnamefont {Hillery}}, \ and\ \bibinfo {author}
  {\bibfnamefont {M.}~\bibnamefont {Orszag}},\ }\href@noop {} {\bibfield
  {journal} {\bibinfo  {journal} {Phys. Rev. A}\ }\textbf {\bibinfo {volume}
  {41}},\ \bibinfo {pages} {2756} (\bibinfo {year} {1990})}\BibitemShut
  {NoStop}%
\bibitem [{\citenamefont {Plimak}\ \emph {et~al.}(2001)\citenamefont {Plimak},
  \citenamefont {Olsen}, \citenamefont {Fleischhauer},\ and\ \citenamefont
  {Collett}}]{PLIMAK}%
  \BibitemOpen
  \bibfield  {author} {\bibinfo {author} {\bibfnamefont {L.~I.}\ \bibnamefont
  {Plimak}}, \bibinfo {author} {\bibfnamefont {M.~K.}\ \bibnamefont {Olsen}},
  \bibinfo {author} {\bibfnamefont {M.}~\bibnamefont {Fleischhauer}}, \ and\
  \bibinfo {author} {\bibfnamefont {M.~J.}\ \bibnamefont {Collett}},\ }\href
  {http://stacks.iop.org/0295-5075/56/i=3/a=372} {\bibfield  {journal}
  {\bibinfo  {journal} {EPL (Europhysics Letters)}\ }\textbf {\bibinfo {volume}
  {56}},\ \bibinfo {pages} {372} (\bibinfo {year} {2001})}\BibitemShut
  {NoStop}%
\bibitem [{\citenamefont {Polkovnikov}(2003)}]{Polkovnikov}%
  \BibitemOpen
  \bibfield  {author} {\bibinfo {author} {\bibfnamefont {A.}~\bibnamefont
  {Polkovnikov}},\ }\href {\doibase 10.1103/PhysRevA.68.053604} {\bibfield
  {journal} {\bibinfo  {journal} {Phys. Rev. A}\ }\textbf {\bibinfo {volume}
  {68}},\ \bibinfo {pages} {053604} (\bibinfo {year} {2003})}\BibitemShut
  {NoStop}%
\bibitem [{\citenamefont {Vogel}\ and\ \citenamefont
  {Risken}(1989)}]{VogelRisken}%
  \BibitemOpen
  \bibfield  {author} {\bibinfo {author} {\bibfnamefont {K.}~\bibnamefont
  {Vogel}}\ and\ \bibinfo {author} {\bibfnamefont {H.}~\bibnamefont {Risken}},\
  }\href {\doibase 10.1103/PhysRevA.39.4675} {\bibfield  {journal} {\bibinfo
  {journal} {Phys. Rev. A}\ }\textbf {\bibinfo {volume} {39}},\ \bibinfo
  {pages} {4675} (\bibinfo {year} {1989})}\BibitemShut {NoStop}%
\bibitem [{\citenamefont {Nyman}\ and\ \citenamefont
  {Szyma\ifmmode~\acute{n}\else \'{n}\fi{}ska}(2014)}]{Nyman}%
  \BibitemOpen
  \bibfield  {author} {\bibinfo {author} {\bibfnamefont {R.~A.}\ \bibnamefont
  {Nyman}}\ and\ \bibinfo {author} {\bibfnamefont {M.~H.}\ \bibnamefont
  {Szyma\ifmmode~\acute{n}\else \'{n}\fi{}ska}},\ }\href@noop {} {\bibfield
  {journal} {\bibinfo  {journal} {Phys. Rev. A}\ }\textbf {\bibinfo {volume}
  {89}},\ \bibinfo {pages} {033844} (\bibinfo {year} {2014})}\BibitemShut
  {NoStop}%
\bibitem [{\citenamefont {Pitaevskii}\ and\ \citenamefont
  {Stringari}(2004)}]{PITA2004}%
  \BibitemOpen
  \bibfield  {author} {\bibinfo {author} {\bibfnamefont {L.}~\bibnamefont
  {Pitaevskii}}\ and\ \bibinfo {author} {\bibfnamefont {S.}~\bibnamefont
  {Stringari}},\ }\href@noop {} {\emph {\bibinfo {title} {Bose Einstein
  condensation}}}\ (\bibinfo  {publisher} {Clarendon Press, Oxford},\ \bibinfo
  {year} {2004})\BibitemShut {NoStop}%
\end{thebibliography}%

\end{document}